\documentclass[fleqn,usenatbib]{mnras}

\usepackage{newtxtext,newtxmath}
\usepackage[T1]{fontenc}
\usepackage{ae,aecompl}

\usepackage{graphicx}	% Including figure files
\usepackage{amsmath}	% Advanced maths commands
\usepackage{amssymb}	% Extra maths symbols

\usepackage{subfig}

\usepackage{booktabs}

\usepackage{cleveref}
\crefname{section}{§}{§§}
\Crefname{section}{§}{§§}

\newcommand{\TTO}{T$_{\rm 21}$}

\newcommand{\Lya}{Ly-${\alpha}$}

\newcommand{\Vc}{$V_{c}$}

   \title[Radio galaxies: 21-cm fluctuations]{High-redshift radio galaxies: a potential new source of 21-cm fluctuations}
\author[Reis et al.]{
Itamar Reis$^{1}$\thanks{E-mail: itamarreis@mail.tau.ac.il},
Anastasia Fialkov$^{2}$,
and Rennan Barkana$^{1}$
\\
$^{1}$School of Physics and Astronomy, Tel-Aviv University, Tel-Aviv, 69978, Israel\\
$^{2}$Institute of Astronomy, University of Cambridge, Madingley Road, Cambridge CB3 0HA, UK\\
}

\date{Accepted XXX. Received YYY; in original form ZZZ}

\pubyear{2020}

\begin{document}
\label{firstpage}
\pagerange{\pageref{firstpage}--\pageref{lastpage}}
\maketitle

\begin{abstract}
Radio sources are expected to have formed at high redshifts, producing an excess radiation background above the cosmic microwave background (CMB) at low frequencies. Their effect on the redshifted 21-cm signal of neutral hydrogen is usually neglected, as it is assumed that the associated background is small. Recently, an excess radio background above the level of the CMB has been proposed as one of the possible explanations for the unusually strong 21-cm signal from redshift $z\sim 17$ reported by the EDGES collaboration. As a result, the implications of a smooth and extremely strong excess radio background on both the sky-averaged (global) 21-cm signal and its fluctuations have been considered. Here we take into account the inhomogeneity of the radio background created by a population of high-redshift galaxies, and show that it adds a new type of 21-cm fluctuations to the well-known contributions of density, velocity, \Lya{} coupling, heating and reionization. We find that a population of high-redshift galaxies even with a moderately-enhanced radio efficiency (unrelated to the EDGES result) can have a significant effect on the 21-cm power spectrum and global signal in models with weak X-ray heating. For models that can explain the EDGES data, we conduct a large parameter survey to explore their signatures. We show that in such models the 21-cm power spectrum at $z\sim 17$ is enhanced by up to two orders of magnitude compared to the CMB-only standard case, and the shape and time evolution of the power spectrum is significantly modified by the radio fluctuations. These fluctuations are within reach of upcoming radio interferometers. We also find that these models can be significantly constrained by current and future observations of radio sources.
\end{abstract}


%%%%%%%%%%%%%%%%%%%%%%%%%%%%%%%%%%%%%%%%%%%%%%%%%%

%%%%%%%%%%%%%%%%% BODY OF PAPER %%%%%%%%%%%%%%%%%%

\section{Introduction}

The faint 21-cm signal of high-redshift neutral hydrogen, made visible against the background radiation by the light of the first stars and galaxies, is currently one of the major targets of observational astronomy. The first tentative detection of the sky-averaged (global) signal from cosmic dawn, an epoch roughly corresponding to the redshift range of $z\sim 15-27$,  was recently reported by the EDGES collaboration \citep{bowman18}. With the true nature of this signal still  being debated \citep[e.g.,][]{Hills:2018, Bradley:2019, Singh:2019, Spinelli:2019, Sims:2020}, fast advances are being made to validate this detection using radiometers such as the Large-Aperture Experiment to Detect the Dark Ages \citep[LEDA,][]{bernardi16, Price:2018}, the Shaped Antenna measurement of the background RAdio Spectrum \citep[SARAS,][]{patra13,Singh:2018}, Probing Radio Intensity at high-Z \citep[PRIZM,][]{philip19}, the Mapper of the IGM Spin Temperature\footnote{http://www.physics.mcgill.ca/mist/} (MIST), and the Radio Experiment for the Analysis of Cosmic Hydrogen\footnote{https://www.kicc.cam.ac.uk/projects/reach} (REACH), and interferometers such as the low band of the Low Frequency Array \citep[LOFAR,][]{Gehlot:2019}, the Hydrogen Epoch of Reionization Array \citep[HERA,][]{deboer17} and the Square Kilometer Array (SKA) \citep[SKA,][]{koopmans15}.

The hydrogen radio signal from the Epoch of Reionization \citep[EoR, $z\sim 6-15$, e.g.,][]{planckcollaboration18} and cosmic dawn is produced by atoms located in the intergalactic medium (IGM) at the rest-frame 1.42 GHz frequency, corresponding to a wavelength of 21 cm. Due to the expansion of the Universe, the signal redshifts and can be detected by radio telescopes at frequencies below 200 MHz against the background radiation, which is usually assumed to be the cosmic microwave background (CMB). The 21-cm signal is sensitive to the thermal and ionization states of the IGM and, thus,  probes the formation of the first stars and galaxies through their effect on the gas \citep[e.g., see][]{barkana18book, Mesingerbook}. As the first sources of light turn on, their radiation impacts the 21-cm signal through several processes. Perhaps the most important effect, owing to which the 21-cm signal becomes distinguishable from the background radiation, is coupling of the hydrogen spin temperature (the effective temperature of the 21cm transition) to the kinetic gas temperature by stellar  \Lya{} photons \citep[the Wouthuysen-Field, or WF,  effect,][]{wouthuysen52, field58}.  Because at the dawn of star formation  the thermal evolution of the IGM is dominated by adiabatic cooling,  the gas  temperature is lower than that of the  background and the resulting 21-cm signal is seen in absorption. As the first population of astrophysical sources evolves, X-ray binaries form and contribute to the heating budget, driving the temperature of the gas up. Depending on its efficiency, the astrophysical heating can result in either an absorption or emission 21-cm signal at the EoR redshifts \citep{fialkov14a}. Finally, the process of reionization takes place, in which ultraviolet (UV) photons produced by stars ionize the intergalactic hydrogen and eliminate the 21-cm signal from the IGM. 

Out of the outlined scenario, the strongest constraints are on the EoR, owing to it being more accessible observationally. Galaxy surveys provide insight on the process of star formation and constrain properties of galaxies at $z\lesssim 10$ \citep[e.g., see][ and references therein]{Behroozi:2019}; meanwhile, the ionization history is constrained by observations of high redshift quasars and galaxies that suggest that the Universe was largely neutral at $z\gtrsim 7.5$ \citep[e.g.,][]{Mason:2018, Banados:2018} in agreement with measurements of the CMB temperature and polarization by the {\it Planck} satellite  \citep[e.g.,][]{planckcollaboration18}. 

Radio telescopes have begun to contribute to the understanding of cosmic dawn and the EoR. The tentative EDGES detection, if true, requires efficient star formation at $z\sim 20$ \citep{fialkov19, mirocha19, Schauer:2019} and, thus, could be the first real evidence for the first population of stars.  Other high-redshift experiments are publishing upper limits which, however,  are still too weak to significantly constrain theoretical models \citep{bernardi16, eastwood19, Gehlot:2019}. At lower redshifts associated with the EoR, non-detections of the global signal have been reported, ruling out possible astrophysical scenarios that include a cold IGM and/or abrupt reionization \citep{monsalve17,Monsalve:2018, monsalve19, Singh:2018}. At the same time, upper limits on the fluctuations in the signal are being set using interferometers such as LOFAR \citep{LOFAR-EoR:2020}, which has recently allowed us to constrain the thermal and ionization properties of the IGM as well as the amplitude of the radio background at $z=9.1$ \citep{Ghara:2020, Mondal:2020}; other interferometers include the Precision Array to Probe the Epoch of Reionization \citep[PAPER,][]{Kolopanis:2019}, the Murchison Wide-field Array \citep[MWA,][]{Barry:2019, Trott:2020}, and the GMRT-EoR experiment \citep[][]{paciga11}. Meanwhile, arrays under construction, including HERA, the SKA, and the New Extension in Nancay Upgrading LOFAR \citep[NenuFAR,][]{zarka12}, promise to provide accurate measurements of the fluctuations from a wide range of redshifts and scales. 

The recent EDGES detection \citep{bowman18} has motivated new directions for theoretical study. There are two plausible categories of explanations of the anomalously deep signal of $-500^{+200}_{-500}$ mK at $z\sim 17$. One involves a baryon - dark matter interaction that can produce faster than adiabatic cooling of the gas \citep{barkana18, Berlin, barkana18a, munoz18, Liu19}. The other category, which we focus on in this paper, involves an excess radio background at high redshifts above the level of the CMB \citep{bowman18, feng18, ewall18, fialkov19, mirocha19, ewall20}. High-redshift sources,  such  as  radio loud active galactic nuclei  \citep[AGN,][]{urry95, biermann14, bolgar18, ewall18, ewall20} or star-forming galaxies \citep{condon92, jana19},  could contribute to the radio background, thus affecting the 21-cm signal. The properties of typical high-redshift radio galaxies are poorly understood due to the lack of sensitive observations. Bright radio galaxies are seen out to $z\sim5$ and are extremely rare \citep{miley08}, while the unresolved population might be contributing to the diffuse radio background as suggested by \citet{bridle67}. The excess of the radio background above the CMB level,  observed by ARCADE2 \citep{fixsen11, seiffert11} and confirmed by LWA1 \citep{dowell18}, could partially be explained by a population of high-redshift unresolved sources, although it is still highly uncertain what fraction of the excess is of extragalactic origin \citep[e.g., see][]{Subrahmanyan:2013}.  \citet{dowell18} showed that the observed excess has a spectral index of $-2.58$ and a brightness temperature of $\sim 603$ K at the rest-frame 21-cm frequency. 

Assuming a phenomenological parameterization of a high-redshift radio excess with a synchrotron spectrum, \citet{fialkov19} showed that it could explain the EDGES signal even if it made up  only a small fraction (half a percent) of the excess observed by ARCADE2 and LWA1.  Such a contribution   could be created by {\it exotic} agents,  e.g., annihilating dark matter or super-conducting cosmic strings \citep{Fraser:2018, Pospelov:2018, Brandenberger:2019}. Such models are generally less strongly constrained by other observations than the dark matter cooling models from the alternative class of explanations. Still, astrophysical radio sources would provide a much more natural explanation of an early radio excess. However, all proposed models face challenges to create such a strong excess radio emission as required by EDGES. For example, \citet{mirocha19} assumed a ratio emissivity proportional to the star formation rate (SFR), as observed for present-day galaxies \citep{gurkan18}, and found that high redshift galaxies need to be $1000$ times  brighter in radio than their present-day counterparts in order to explain the EDGES signal. \citet{ewall20} explored the possibility of radio-loud accretion onto black holes. One of the difficulties of this model is the expected suppression of synchrotron emission from relativistic electrons due to inverse Compton scattering of these electrons off the CMB, which is stronger at high redshifts than today owing to the higher energy density of the CMB \citep{ghisellini14, saxena17}. \citet{jana19} considered radio emission from supernova explosions of massive population III stars. However, this model also predicts  the production of cosmic rays which can heat the IGM and significantly counteract the effect of the enhanced radio background on the 21-cm signal.

All the above-mentioned studies focused on the implications of an anomalously strong and {\it smooth}   excess radio background on the global 21-cm signal, with the exception of \citet{fialkov19} who considered the impact of a smooth background on the 21-cm fluctuations but also derived an analytical formula to include the radio background fluctuations in the linear regime. In this work we explore the 21-cm implications of an excess radio background produced by galactic halos, accounting fully for its non-linear spatial inhomogeneity. We focus on three main points: (1) assuming that the radio emission scales with the SFR of each galactic halo, we use semi-numerical methods to explore the impact of a non-uniform radio background on both the global 21-cm signal and the fluctuations; (2) we quantify the contribution of a standard population (i.e., not normalized to explain EDGES) of radio galaxies to both the global 21-cm signal and the fluctuations; and (3) we conduct a large parameter study to investigate the implications of the EDGES-motivated scenarios for the 21-cm power spectra.

This paper is organized as follows: In \cref{sec:methods} we briefly review our 21-cm simulation. In \cref{sec:radio_back} we describe the framework used to include the excess radio background in the simulation and summarize the  effects of the excess radio background on the 21-cm signal. In \cref{sec:case_study} we explore the signature of  the radio background fluctuations on the 21-cm signal. In \cref{sec:weak_radio} we consider a standard population of high-redshift radio galaxies (too weak to explain EDGES), and explore their signature in the 21-cm signal. In \cref{sec:parameters} we perform a parameter study, looking at the range of possible 21-cm power spectra in model that fit the EDGES global signal result and constrain model parameters using the EDGES detection.   We summarize in \cref{sec:summary}. Throughout this paper we adopt cosmological parameters from \citet{Planck:2014}. All scales and wavenumbers are in comoving units.

\section{Methods}
\label{sec:methods}

We use our own  semi-numerical code to calculate the 21-cm signal \citep[e.g.,][]{visbal12, fialkov14, cohen17, fialkov19}. The simulation computes all the required ingredients necessary for the evaluation of the 21-cm signal, such as  the  density and kinetic temperature of the gas, the coupling coefficient, spin temperature and the neutral fraction. All these quantities are calculated on a 128$^3$ grid (for the purposes of this work) with a resolution of 3 comoving Mpc. At the output of the simulation we obtain cubes of the 21-cm brightness temperature and calculate the global signal and its spherically averaged power spectrum at every redshift. The code was originally inspired by the architecture of {\sc 21cmFAST} \citep{mesinger11}, but has a completely independent implementation and some different ingredients.

The initial conditions generator receives as inputs power spectra of density and velocity fields \citep[calculated using the publicly available code  {\sc CAMB},][]{camb} and outputs three-dimensional cubes of density fluctuations and relative velocity between dark matter and baryons \citep{tseliakhovich10}. The density and velocity fields are evolved using linear perturbation theory. The number of dark matter halos in each cell of 3$^3$ comoving Mpc$^3$ is determined based on the values of the local density and relative velocity and is derived at each redshift using a modified Press-Schechter model \citep{press74, sheth99, barkana04}.

Galaxies are assumed to form in halos with circular velocity higher than a minimum threshold, \Vc{}, in which stars form with a star formation efficiency $f_{*}$.
The circular velocity is related to the minimum mass $M$ of dark matter halo in which stars can form through
\begin{equation}
    V_{c} = 16.9 \left(\frac{M}{10^8 M_{\odot}}\right)^{1/3}\left(\frac{1+z}{10}\right)^{1/2}  \left(\frac{\Omega_m h^2}{0.141}\right)^{1/6} \left(\frac{\Delta_c}{18 \pi^2}\right)^{1/6}\   {\rm km}\;{\rm s}^{-1}\ ,
\end{equation}
where $\Delta_c$ is the ratio between the collapsed  density and the critical density at the time of collapse, and equals $18 \pi^2$ for spherical collapse. 
In addition to this cutoff, we also implement suppression in star formation (realized as an environment-dependent boost in the minimum mass of star forming halos) due to the relative velocity between dark matter and baryons \citep{tseliakhovich10,fialkov12}, Lyman-Werner feedback \citep{haiman97,fialkov2013}, and photoheating feedback \citep{rees86, sobacchi13,cohen16}. The efficiency $f_*$ is assumed to be constant for halos above the atomic cooling threshold corresponding to $V_c = 16.5$ km s$^{-1}$, and features a logarithmic suppression in halos down to the molecular cooling threshold of $V_c = 4.2$   km s$^{-1}$ \citep[for details see ][]{cohen19}.

Given a population of galaxies we calculate the radiation fields that affect the IGM and, thus, the 21-cm signal.  The intensity of the \Lya{} radiation, $J_{\alpha}$, is calculated assuming galaxies contain population II stars (the modification of radiation intensities that corresponds to population III stars is small compared to the wide range of astrophysical parameters that we consider here).  $J_{\alpha}$ regulates the strength of the WF coupling and also contributes to the heating of the IGM (see below). The X-ray luminosity of galactic halos is assumed to scale with their star formation rate, as is the case for observed galaxies \citep[e.g.,][]{Mineo:2012, fragos13, fialkov14a, Pacucci:2014}, with an efficiency factor $f_{X}$, where $f_X = 1$ corresponds to the X-ray luminosity of present-day galaxies located in low-metallicity regions. For the majority of this work  we assume that the X-ray spectral energy distribution (SED) has a power law shape with a slope $\alpha$ and a low-frequency cutoff $\nu_{\rm min}$. However, we also make use of a realistic SED  derived by \citet{fragos13} for a population of high-redshift X-ray binaries. This SED peaks at energies of $\sim 3$ keV, resulting in a relatively hard X-ray emission. Its effects on the 21-cm signal (compared to a softer power-law SED) have been discussed by \citet{fialkov14a}. Reionization is implemented using the excursion set formalism \citep{furlanetto04}, in which a region is assumed to be ionized if the collapsed fraction exceeds a  $\zeta^{-1}$ threshold, where  $\zeta$ is the overall ionizing efficiency of sources. In our models there is one-to-one correspondence between $\zeta$ and the CMB optical depth, $\tau$ \citep[when other parameters are fixed, see discussion in][]{cohen19}, and higher values of $\zeta$ at higher $V_c$  are needed to give the same $\tau$. Because $\tau$ (rather than $\zeta$) is constrained observationally \citep[e.g., by the {\it Planck} satellite][]{planckcollaboration18}, we usually choose to work with $\tau$. The ionizing photons are assumed to travel up to a maximum distance set by the mean free path, R$_{\rm mfp}$ \citep[e.g.,][]{greig15}. Finally, an excess radio background above the CMB temperature can be present, and its  impact on the 21-cm signal is reviewed in the next section.  \citet{fialkov19} considered a homogeneous {\it external} radio background model (i.e., not constrained to be directly related to the astrophysical sources) with a synchrotron spectrum of spectral index $\beta = -2.6$ and amplitude $A_{\rm r}$ relatively to the CMB at the reference frequency of 78 MHz (the central frequency of the absorption trough detected by EDGES).  In such a model, the excess  radio background over the CMB at redshift $z$ and at the intrinsic 21-cm frequency of 1.42 GHz is given by 
\begin{equation}
T_{\rm Radio} = 2.725 (1+z) \times \left[\frac{1420}{78 (1+z)}\right]^{\beta} A_{\rm r} ~{\rm K}\ ,
\label{Eq:ar}
\end{equation}
where $2.725$ K is the CMB temperature today.
Our model, thus, is based on eight free parameters: $f_*$,  \Vc{},  $f_{X}$, $\alpha$, $\nu_{\rm min}$, $\tau$ (or $\zeta$),  R$_{\rm mfp}$ and the amplitude of the radio background. See \citet{cohen19} and \citet{fialkov19} for more details on the semi-numerical simulation. 

The interplay between different radiative fields in their impact on the 21-cm signal is non-trivial owing to the fact that the different fields can produce 21-cm fluctuations at different scales. In addition, the contribution of these different  sources of 21-cm fluctuations can either correlate or anti-correlate. For example,  \Lya{}  radiation acts to deepen the absorption signal via the WF coupling, while (when the gas is colder than the background radiation) X-rays tend to reduce it by heating up the gas. Therefore, the contributions of the \Lya{} and X-ray fields to the 21-cm signal anti-correlate, creating two peaks in the 21-cm power spectrum when seen at a fixed comoving scale  as a function of redshift \citep[e.g., see][ and also Fig.~\ref{fig:power_spectrum_vs_f_radio} below]{cohen18}. The  highest-redshift cosmic dawn peak is typically dominated by  \Lya{} emission and can be located anywhere between $z\sim 13-35$ \citep[e.g., Fig.4 of][]{cohen18}. The presence or absence of the X-ray peak (at $z\lesssim 25$) depends on the properties of the X-ray sources, most importantly on their SED \citep{ fialkov14a, Pacucci:2014, fialkov14, cohen18}. Hard X-ray photons with $\nu_{\rm min}\gtrsim1$ keV have a long mean free path of a few hundred comoving Mpc, washing out X-ray fluctuations. In this case the X-ray peak might not be apparent in the power spectrum. 
On the other hand, soft X-ray sources  (loosely defined as sources with an SED peaking at frequencies below $\nu_{\rm min}\sim1$ keV) are efficient in heating up the gas on smaller scales and thus imprint  the X-ray peak in the power spectrum. At lower redshifts, a reionization peak is typically present with the maximum power at the redshift corresponding to a neutral fraction of $\sim 50$\%. The effect of an enhanced radio background, and how it might modify this picture, are described in the rest of this paper.

In this work we use an upgraded version of our code. Instead of the phenomenological homogeneous external radio background, we consider a {\it galactic} radio background, i.e., a fluctuating field produced by radio galaxies (see \cref{sec:radio_back} for more details). 
Moreover, compared to the previous version, here we include a number of new physical  processes that make our simulation more realistic: 
\begin{enumerate}
    \item We use a Poisson realization of the mean number of halos in each pixel (more accurately, the mean number of halos created in each time step), instead of the mean number itself. Poisson fluctuations in the spatial distribution of galaxies are important when the radiation backgrounds are dominated by a relatively small number of rare galaxies.
    \item We include the effect of multiple scattering of \Lya{} photons on their spatial distribution. While previously \Lya{} photons were assumed to travel in straight lines between emission and absorption, here we take into account the fact that these photons can scatter off the wings of the \Lya{} line, before being absorbed at the line center. This results in photons traveling smaller effective distances from their sources and enhances the spatial fluctuations in  $J_{\alpha}$, and thus fluctuations in the coupling field and in the 21-cm signal \citep{Chuzhoy_2007,semelin07,naoz08}.
    \item The heating of the IGM by \Lya{} photons is also taken into account. This heating mechanism is significant in models with highly inefficient X-ray heating, i.e., with very low $f_{X}$ \citep[e.g.,][]{chen04,chuzhoy06,ghara19}.
\end{enumerate}
We discuss these effects in more detail in \citet{reis20a, reis20b}.

\section{The excess radio background}
\label{sec:radio_back}

An excess radio background  enhances  the contrast between the spin temperature and the background radiation temperature, decreases the strength of the coupling between the spin temperature and the kinetic gas temperature \citep[e.g.,][]{feng18}, and enhances the radiative heating \citep[e.g.,][]{fialkov19}.   

The brightness temperature of the 21-cm signal, \TTO{}, is determined by the contrast between the neutral hydrogen spin temperature, $T_{\rm S}$, and the temperature of the background radiation field, $T_{\rm rad}$:
\begin{equation}
    T_{21} = \frac{T_{\rm S} - T_{\rm rad}}{1+z}(1 - e^{-\tau_{21}})\ . \label{eq:T21}
\end{equation}
Here $\tau_{21}$ is the 21-cm optical depth 
\begin{equation}
\tau_{21} = \frac{3h_{\rm pl}A_{10}c \lambda_{21}^2n_{\rm H}}{32\pi k_{B} T_{\rm S} (1+z) dv/dr}\ ,
\end{equation}
where  $dv/dr = H(z)/(1+z)$ is the gradient of the line of sight component of the comoving velocity field, $H(z)$ is the Hubble rate, $A_{10}$ is the spontaneous decay rate of the hyper-fine transition, and $n_{\rm H}$ is the neutral hydrogen number density which depends on the ionization history driven by both ultraviolet and X-ray photons. The rest of the parameters assume their standard definitions.
A useful approximation for the 21-cm brightness temperature in the low optical depth regime ($\tau_{21} \ll 1$) is given by
\begin{multline}
\label{eqn:tb}
T_{\rm 21}  \approx  26.8 \left(\frac{\Omega_{\rm b} h}{0.0327}\right)
\left(\frac{\Omega_{\rm m}}{0.307}\right)^{-1/2}
\left(\frac{1 + z}{10}\right)^{1/2} \\
(1 + \delta) x_{\rm HI} \frac{x_{\rm tot}}{1 + x_{\rm tot}} \left( 1 - \frac{T_{\rm rad}}{T_{\rm K}}\right) ~{\rm mK}\ .
\end{multline}

The background radiation is usually assumed to be the CMB with $T_{\rm CMB} = 2.725(1+z)$ K. However, in the presence of an extra radio contribution of brightness temperature  $T_{\rm Radio}$ the total background is 
\begin{equation}
    T_{\rm rad} = T_{\rm CMB} + T_{\rm Radio}\ .
\end{equation}
A more subtle effect of $T_{\rm Radio}$ is on $T_{\rm S}$ through its effect on the coupling coefficients between the spin temperature and  the kinetic temperature of the hydrogen gas \citep{feng18, barkana18book, fialkov19}:
\begin{equation}
\label{eqn:xalpha}
    x_{\alpha} = \frac{1}{A_{10} T_{\rm rad}} \frac{16 \pi^2 T_{*} e^2 f_{\alpha}}{27 m_{e} c}  J_{\alpha}\ ,
\end{equation}
and 
\begin{equation}
\label{eqn:col}
    x_c = \frac{1}{A_{10} T_{\rm rad}} \kappa_{1-0}(T_{\rm K}) n_{\rm H} T_{\star}\ .
\end{equation}
Here $T_{*} = 0.0682$ K, $f_{\alpha} = 0.4162$ is the oscillator strength of the \Lya{} transition, and $\kappa_{1-0}(T_{\rm K})$ is a known atomic coefficient \citep{allison69, zygelman05}. Given these coupling coefficients, the spin temperature is
\begin{equation}
T_{S} = \frac{x_{\rm rad} + x_{\rm tot}}{x_{\rm rad} T_{\rm rad}^{-1} + x_{\rm tot} T_{\rm K}^{-1}}\ ,
\end{equation}
where 
\begin{equation}
x_{\rm tot} = x_{\alpha} + x_c\ , 
\label{eq:coupling}
\end{equation}
and
\begin{equation}
    x_{\rm rad} = \frac{1 - e^{-\tau_{21}}}{\tau_{21}}
\end{equation}
is the radiative coupling \citep{venumadhav18}. Finally, the radio background also affects the value of the kinetic temperature, based on the CMB heating recently noted by \citet{venumadhav18}. The corresponding heating rate is
\begin{equation}
\label{eqn:epsilon_rad}
    \epsilon_{\rm rad} = \frac{x_{\rm HI} A_{10} }{2 H(z)} x_{\rm rad} \left(\frac{T_{\rm rad}}{T_{\rm S}} - 1 \right) \frac{T_{21}}{T_{\rm K}}\ ,
\end{equation}
 where $x_{\rm HI}$ is the neutral fraction. 

In addition to the overall enhancement of the signal (when it is in absorption), the leading order effect of an excess radio background is to, effectively, slow down the cosmic clock. In the presence of an extra radio background it takes longer for the coupling and heating terms  to saturate. The  growth of the \Lya{} coupling term (Eqs. \ref{eqn:xalpha} and \ref{eq:coupling}) is suppressed by a factor $1/T_{\rm rad}$. Compared to the case where the background radiation is the CMB and where \Lya{} coupling is saturated by $z\sim 15$ \citep[a wide range of models was explored by ][see their Fig. 1]{cohen18}, in the presence of a strong radio background the coupling term remains important even at the reionization redshifts. Moreover, in models where the radio background scales with the SFR, as considered in this paper, the coupling effect on $T_{21}$ (approximately $x_{\alpha}/(1+x_{\alpha})$ once the collisional coupling is negligible, if $\tau_{21}$ is small) never saturates to unity, but converges to a different parameter-dependent constant.  This is because $x_\alpha$ depends on the ratio $J_\alpha/T_{\rm rad}$, with both terms proportional to the SFR. Similar reasoning applies to the effect of the gas temperature on the 21-cm signal: higher gas temperatures are needed to saturate the heating term ($1-T_{\rm rad}/T_{\rm K}$ in Eq.~\ref{eqn:tb}), and even before that, the moment of the heating transition (defined as the redshift at which the gas heats up to the temperature of the background radiation) is shifted to lower redshifts. As a result,  the global 21-cm signal is mostly seen in absorption, with the emission feature either reduced to a small redshift range or not present at all.  Also, with a stronger radio background, the heating fluctuations do not saturate and can affect the shape of the 21-cm power spectrum at a much wider range of redshifts than in the CMB-only case. 
Therefore, an enhanced radio background can have important implications for  the 21-cm signal from cosmic dawn and the EoR - the redshift range targeted by radiometers and interferometers\footnote{We show an example of the effect of the radio background on the coupling and heating terms in Appendix \ref{sec:radio_general_effect_example}.}.

We note that throughout this section we have assumed a single value of $T_{\rm rad}$ for each pixel. In reality there are two additional effects that we neglect. First, $T_{21}$ in each pixel is effectively measured with respect to the sky-averaged (cosmic mean) value $T_{\rm rad}$; our use of the local $T_{\rm rad}$ for this in eq.~(\ref{eq:T21}) corresponds to neglecting the direct effect of fluctuations in the radio background, which would contribute even in the $\tau_{21} = 0$ limit and thus may be seen as a source of additional noise for 21-cm measurements (but could also be interesting in itself). Second, 21-cm absorption occurs along the line of sight, so the anisotropy of the distribution of radio sources around each pixel adds another contribution. We leave for further work a detailed exploration of these additional effects.

\subsection{Radio background from galaxies}
\label{sec:radiobck}

The novelty of this work is in considering fluctuations in the radio background and evaluating their effect on the 21-cm signal. We assume the galaxy radio luminosity per unit frequency, calculated in units of W Hz$^{-1}$, to be proportional to the star formation rate (SFR),
\begin{multline}
\label{eqn:eps_radio}
    L_{\rm Radio} (\nu, z ) = f_{\rm Radio} 10^{22} \left(\frac{\nu}{150 {\rm MHz}}\right)^{-\alpha_{\rm Radio}} \frac{\rm SFR}{M_{\odot} \rm{yr}^{-1} }\ ,
\end{multline}
based on the empirical relation of \citet{gurkan18} and similarly to \citet{mirocha19}.   In Eq.~\ref{eqn:eps_radio}, $\alpha_{\rm Radio}$ is the spectral index in the radio band, which we set to $\alpha_{\rm Radio} = 0.7$ as in \citet{mirocha19, gurkan18}, see also \citet{condon02, heesen14}.   $f_{\rm Radio}$ is the normalization of the radio emissivity, where the average value for present day star forming galaxies is $f_{\rm Radio} = 1$; while observations indicate a significant scatter, here we assume a uniform value of $f_{\rm Radio}$ (which we vary over a wide range), and leave for future work a direct consideration of such scatter.

We calculate the temperature of the radio background  at redshift $z$, at the 21-cm frequency $\nu_{21}$ by integrating over the contribution of all galaxies within the past light-cone \citep[following][]{ewall20}:
\begin{multline}
\label{eq:Tgal}
    T_{\rm Radio} (\nu_{21}, z) = \frac{\lambda_{21}^2}{2 k_{\rm B}} \frac{c (1+z)^3}{4 \pi} \\ \int\epsilon_{\rm Radio}\left(\nu_{21} \frac{1 + z_{\rm em}}{1 + z}, z_{\rm em}\right) (1 + z_{\rm em})^{-1} H(z_{\rm em})^{-1} dz_{\rm em}\ ,
\end{multline}
where $z_{\rm em} > z$ is the redshift of emission. The comoving radio emissivity, $\epsilon_{\rm Radio}$, is the luminosity per unit frequency per unit comoving volume, averaged in this spherical integral over radial shells (with a radius corresponding to the light travel distance between $z_{\rm em}$ and $z$). We note that this spherical shell calculation is similar to the calculation of the X-ray and \Lya{} radiation fields in our simulation, except that for X-rays the optical depth of the IGM also plays a role while the radio photons are not absorbed in the IGM, and for \Lya{} we take into account the effect of multiple scattering which requires using different window functions instead of the spherical shells \citep{reis20b}.

Compared to the smooth external radio background (Eq.~\ref{Eq:ar}) which effectively declines with cosmic time, the radio background from galaxies (Eq.~\ref{eq:Tgal}) is non-uniform and increases with time, tracing the growth of galaxies. As we show below,  these differences strongly affect the shape of the 21-cm signal and, thus, the astrophysical parameter constraints implied by the EDGES low-band detection (\cref{sec:parameters}).  

\section{The signature of radio background fluctuations in the 21-cm signal}
\label{sec:case_study}

We first present a general discussion of the various effects of the radio background fluctuations on the 21-cm signal. As we show here, this effect can vary between models, scales and epochs, and can have important implications for the  21-cm signal in the redshift range relevant for the existing and upcoming radio telescopes.

At high redshift, when the 21-cm signal is  driven by the \Lya{} coupling (and, to some extent, by density fluctuations),  the 21-cm fluctuations can be enhanced due to the positive correlation between the contributions from  the radio field and both the coupling  and the density fields. During the coupling transition, individual galaxies produce "coupled bubbles" around them, where the spin temperature is coupled to the kinetic temperature of the gas. This can be seen as an absorption 21-cm feature surrounding individual galaxies. The radio emission from these galaxies enhances the contrast between the spin temperature and the radiation temperature inside the same coupled bubbles,  which results in a stronger 21-cm absorption compared to the case with a uniform radio background of intensity equal to the mean intensity of the fluctuating case. 

\begin{figure*}
    \centering
 \includegraphics[width=0.8\textwidth]{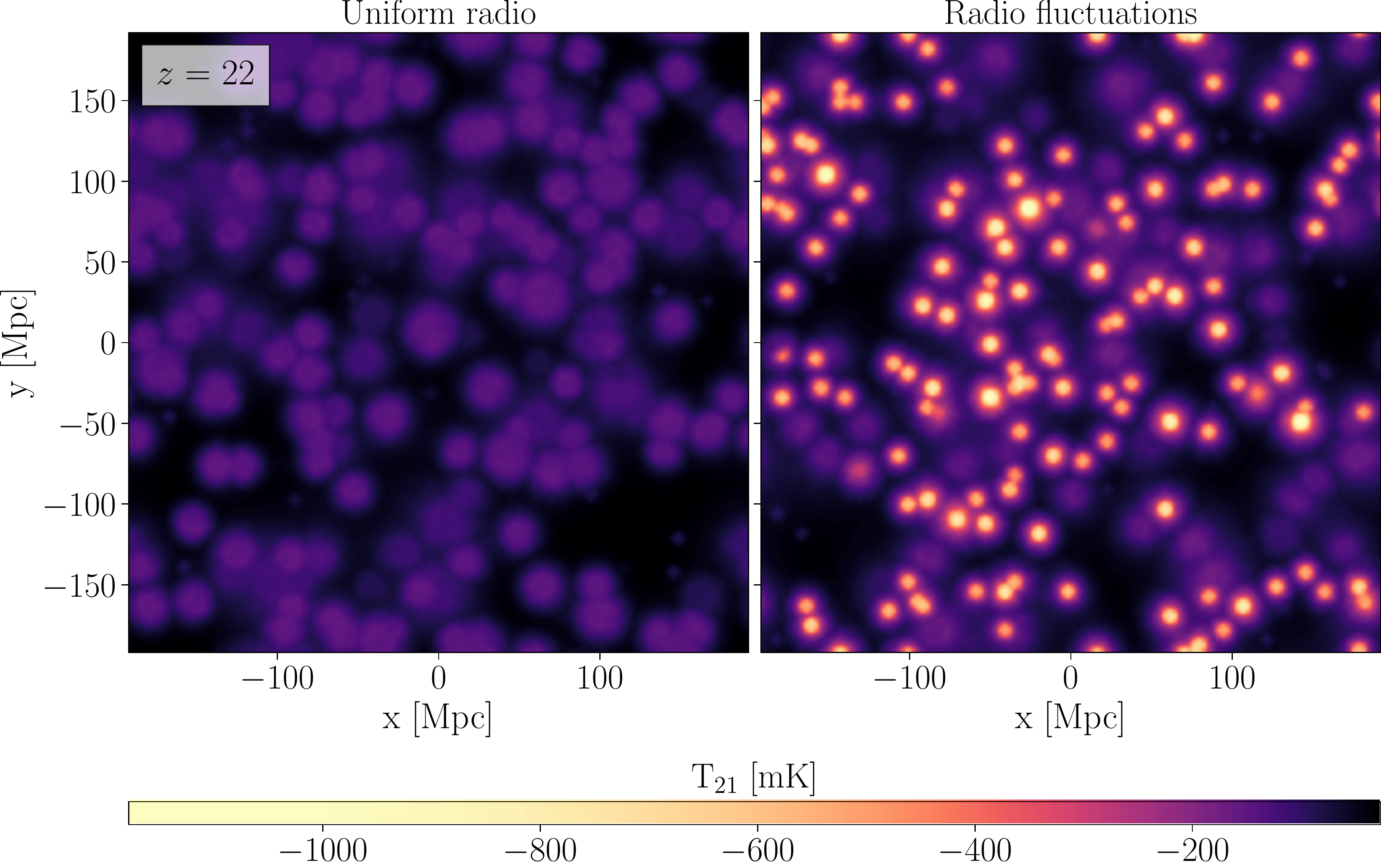} 
    \caption{Effect of fluctuations in the excess radio background on the cosmic dawn 21-cm signal from $z=22$. We compare the case of a uniform radio background  ({\bf left}) to a case where the    radio background is emitted by galaxies ({\bf right}). Both models have the same mean intensity of the radio background. Astrophysical model parameters are:  $V_c = 35$ km s$^{-1}$, $f_* = 0.4$, $f_{\rm Radio} = 1000$.  The X-ray and reionization parameters are irrelevant here, as for this example we focus only on the coupling transition, before the contributions of heating and ionization are significant.
    Each panel  shows the corresponding 21-cm signal  projected in the direction perpendicular to the image. The projected signal is obtained by taking the minimum value along each column of the cubic simulation box.
}
    \label{fig:slices_coupling_massive}
\end{figure*}

An illustration of this is shown in Fig.~\ref{fig:slices_coupling_massive}, where we compare the complete model (right panel) to a reference case with a uniform radio background of the same mean intensity (left panel). In the reference case we can still see the coupled bubbles in the 21-cm signal, but their contrast is greatly enhanced in the full model, in which the radio enhancement is clustered around the galactic halos. To highlight the effect of the radio fluctuations, we have used a simulation with moderately massive halos (minimum circular velocity of 35.5 km/s, corresponding to a minimum halo mass for star formation of $3\times 10^{8} M_{\odot}$ at $z = 20$) along with a high value of the radio production efficiency $f_{\rm Radio} = 1000$.

\begin{figure}
    \centering
 \includegraphics[width=0.49\textwidth]{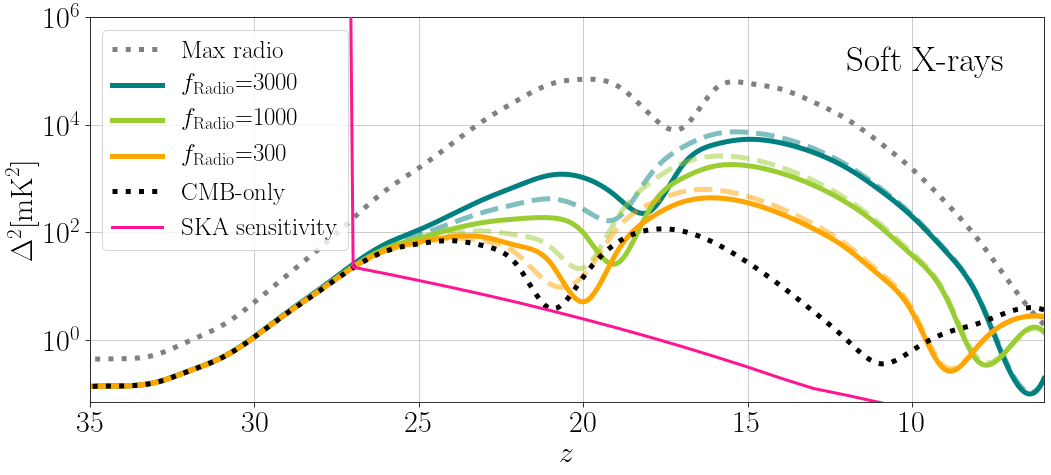} \\  
 \includegraphics[width=0.49\textwidth]{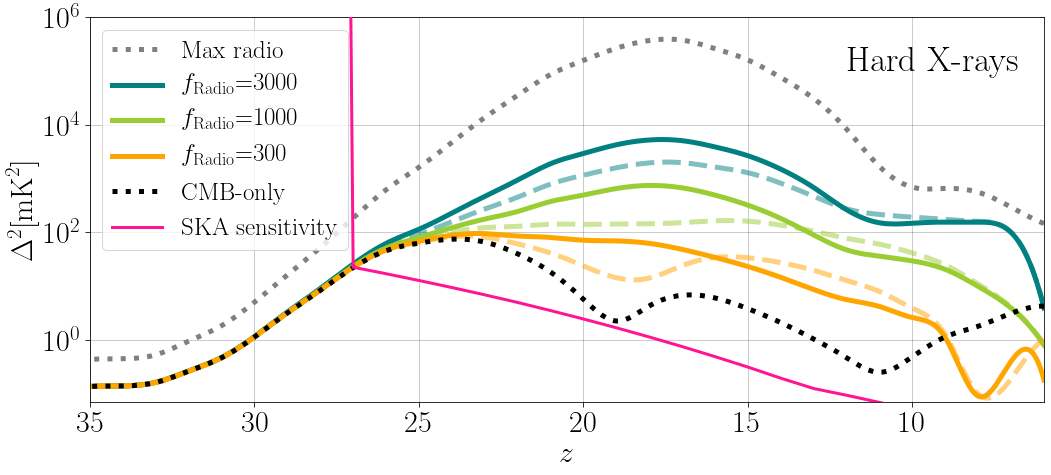}
    \caption{The 21-cm power spectrum at k = 0.1 Mpc$^{-1}$ as a function of redshift for various values of $f_{\rm radio}$ (as indicated in the legend) and for two different X-ray SEDs: soft ($\nu_{\rm min} = 0.1$ keV, {\bf top}) and hard ($\nu_{\rm min} = 1$ keV, {\bf bottom}). The values of the other parameters are fixed: $V_c = 16.5$ km s$^{-1}$, $f_*$ = 0.1, $f_{X}$ = 1, $\alpha = 1$.  We show the full, fluctuating radio background from galaxies (solid) compared to the corresponding smooth radio background with the same mean radio intensity at each redshift as in the fluctuating case (dashed). We also show a case with no excess radio background (i.e., the CMB-only case, black dotted line) and the "maximum radio" limit (see text, grey dotted line). Finally, we also show the SKA1 noise curve (magenta) assuming a single beam, integration time of 1000 hours, 10 MHz bandwidth, and bins of width $\Delta k = k$. }
    \label{fig:power_spectrum_vs_f_radio}
\end{figure}

The effect of the radio background fluctuations on the statistical properties of the 21-cm signal, namely its power spectrum, is shown in Fig.~\ref{fig:power_spectrum_vs_f_radio} for various values of $f_{\rm radio}$ and two different (soft and hard) X-ray SEDs.  Here we choose fairly high values of $f_{\rm Radio}$ to highlight the effects, while models with lower values of $f_{\rm Radio}$ are explored in the next section. We also show two limiting cases: the CMB-only case (i.e., the case with $f_{\rm Radio}=0$) and  the "maximum radio" case. As we can see from Eq.~\ref{eqn:tb}, in the limit  $T_{\rm Radio} \gg T_{\rm CMB}, T_{\rm K}$ the effect of the radio background saturates and the 21-cm brightness temperature becomes independent of $T_{\rm Radio}$:
\begin{equation}
    T_{21} = -26.8 \left(\frac{1 + z}{10}\right)^{1/2} (1+\delta) x_{\rm HI}\,   x_{\rm tot} (T_{\rm CMB}) \frac{T_{\rm CMB}}{T_{\rm K}} {\rm mK}\ ,
\end{equation}
where $x_{\rm tot} (T_{\rm CMB})$ is the coupling coefficient calculated with $T_{\rm rad} = T_{\rm CMB}$ (and we have suppressed the dependence on cosmological parameters). In this limit we neglect the effect of the excess radio background on the heating rate, obtaining a simple upper limit. In the Figure, we compare the effect of a fluctuating radio background to a smooth reference case with the same mean radio intensity at every $z$ as in the fluctuating case.

We find that at high redshifts, prior to the onset of X-ray heating, the inhomogeneous radio background enhances the 21-cm power spectrum by up to a factor of a few, owing to the positive correlation between the contributions of the radio background fluctuations and both the \Lya{} and density fluctuations, and given the negligible contribution of X-ray sources. The boost in power is clearly seen in Fig.~\ref{fig:power_spectrum_vs_f_radio}, where we compare the effect of a fluctuating radio background for each model to a corresponding smooth reference case. We note that Fig.~\ref{fig:slices_coupling_massive} suggests that the effect on 21-cm images of early galaxies may be even more striking than the statistical effect on the power spectrum.

As the universe evolves, heating gradually becomes more important and starts affecting the 21-cm signal. As long as the gas is still colder than the background radiation, heating acts to reduce the 21-cm absorption (by warming up the gas) and, thus, heating fluctuations anti-correlate with 21-cm fluctuations from all the other sources. This usually leads to a local minimum in the 21-cm power spectrum at the epoch when the fluctuations from the various sources nearly cancel out (see also \cref{sec:methods}). In the presence of the radio background fluctuations this local minimum shifts to lower redshifts compared to the smooth reference case (see Fig.~\ref{fig:power_spectrum_vs_f_radio}). Moreover, as we show below, the details of this signature depend strongly on the shape of the X-ray SED.

In the case of a soft X-ray SED, heating fluctuations are  strong enough (compared to the fluctuations imprinted by the  radio background) to produce a clear heating peak  in the power spectrum. However, with radio fluctuations  the height of the heating peak is reduced by  $20 - 30 \%$ compared to the smooth case (with the same mean radio intensity).  When the strength of the radio background is increased, the heating peak is delayed, since the kinetic temperature requires more time to grow before overtaking the enhanced radiation temperature. We note that in this model, the heating peak is seen even in the "maximum radio" limit. 

In comparison, a model with hard X-rays shows a more complicated dependence of the power spectrum on the radio background. In this case heating fluctuations are weaker, and, for a strong enough radio background, never become the dominant source of fluctuations. In such cases the heating peak is not simply delayed or reduced, but is completely washed out, creating in some cases a single overall peak at a redshift that is intermediate between the \Lya{} and heating peaks that would occur in the CMB-only case.

\begin{figure}
    \centering
 \includegraphics[width=0.49\textwidth]{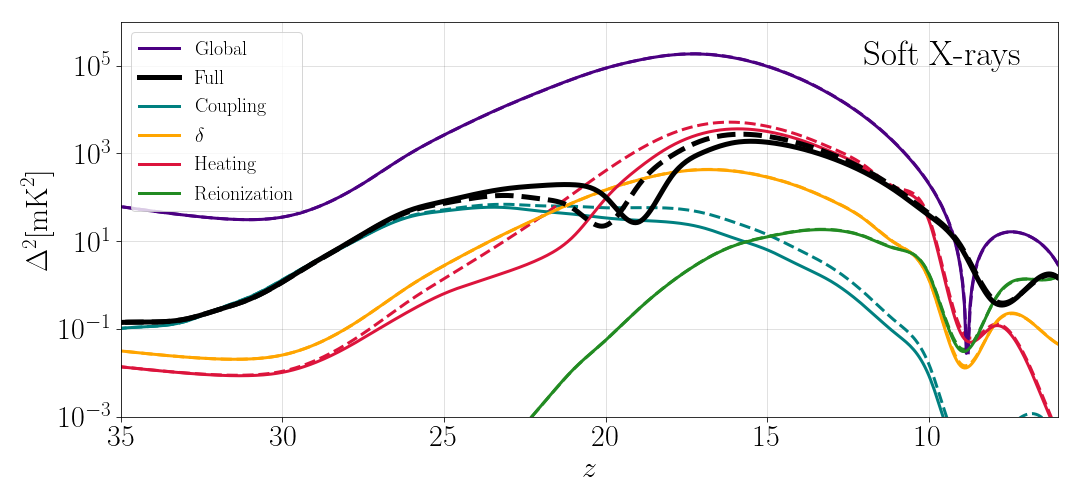}      \\
\includegraphics[width=0.49\textwidth]{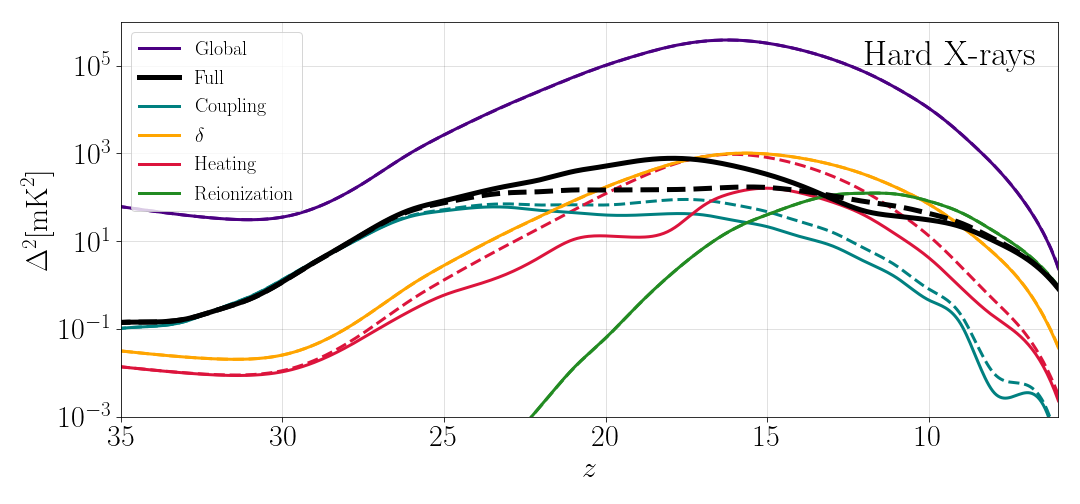}
    \caption{The separate contributions of various sources to the power spectrum at k = 0.1 Mpc$^{-1}$ as a function of redshift  for two different X-ray SEDs: soft ($\nu_{\rm min} = 0.1$ keV, {\bf top}) and hard ($\nu_{\rm min} = 1$ keV, {\bf bottom}).
    We show the coupling term  (blue), the heating term (red), which includes contributions from both the radio and the gas temperature fluctuations, the density term (yellow) and the reionization term (green). The purple line shows the global signal and the black lines show the total power spectrum. We show the same models as in Fig.~\ref{fig:power_spectrum_vs_f_radio}, with $f_{\rm Radio}$ = 1000. We show the models with the fluctuating background (solid lines) or with a uniform radio background of the same mean intensity at each redshift (dashed lines).  
    }
    \label{fig:ps_comps}
\end{figure}

To gain more intuition, we next examine which is the leading
source of fluctuations at every epoch \citep[the same analysis for the CMB-only cases can be found in][for comparison]{cohen18}. We estimate the contribution of each term by allowing variation in the term, while fixing all the other components to their mean values (only here we use the approximation of Eq.~\ref{eqn:tb}, which is linearized in the 21-cm optical depth). The contributions of the density term ($1 + \delta$), the coupling term ($x_{\rm tot}/(1+x_{\rm tot})$), the heating term ($1 - T_{\rm rad}/T_{\rm K}$) and the ionization ($x_{\rm HI}$) to the total 21-cm power spectrum are shown in Fig.~\ref{fig:ps_comps}; note that the contribution of the heating  term takes both the radio and the gas temperature fluctuations into account.  We  explore the importance of each term in models with radio background fluctuations for $f_{\rm Radio} = 1000$ (again, a fairly high value of  $f_{\rm Radio}$ is used to make the effect more apparent) and for hard and soft X-ray SEDs; we again compare to a smooth model with the same mean radio background.

The general behavior in the case with a smooth radio background is similar to what \citet{cohen18} found for the CMB-only case: the signal is dominated by the coupling term at high redshifts, heating and density terms compete at the intermediate redshifts, while  reionization becomes important only at the low-redshift portion of cosmic dawn.
As we see in Fig.~\ref{fig:ps_comps}, adding fluctuations to the radio background strongly affects the signal in the intermediate redshift range, i.e., the redshift range most relevant for the cosmic dawn experiments including the LWA, AARTFAAC Cosmic Explorer,  HERA, NenuFAR and the SKA. Specifically, the strongest effect of the radio fluctuations is on the heating term, $1 - T_{\rm rad}/T_{\rm K}$, through which radio fluctuations and gas temperature fluctuations have opposite effects on the 21-cm signal; the relative balance is SED-dependent as can be clearly seen in Fig.~\ref{fig:ps_comps}. We note that there is also a small but significant effect on the coupling term owing to the competing effects of $J_\alpha$ (in the numerator) and $T_{\rm rad}$ (in the denominator) on $x_\alpha$ in Eq.~\ref{eqn:xalpha}; these two terms are physically correlated, but their opposite effects on $x_\alpha$ lead to reduced 21-cm fluctuations when radio fluctuations are included. 

Fig.~\ref{fig:ps_comps} also shows that during the EoR, 21-cm fluctuations are only weakly affected by radio fluctuations. This is due to the fact that as the radio background rises, and has time to reach larger and larger distances from each source, it becomes increasingly homogeneous. Also the excess radio background does not have a direct effect on the neutral fraction x$_{\rm HI}$, so when fluctuations in x$_{\rm HI}$ dominate 21-cm fluctuations the effect of fluctuations in the excess radio background is reduced. 

\begin{figure}
    \centering
 \includegraphics[width=0.49\textwidth]{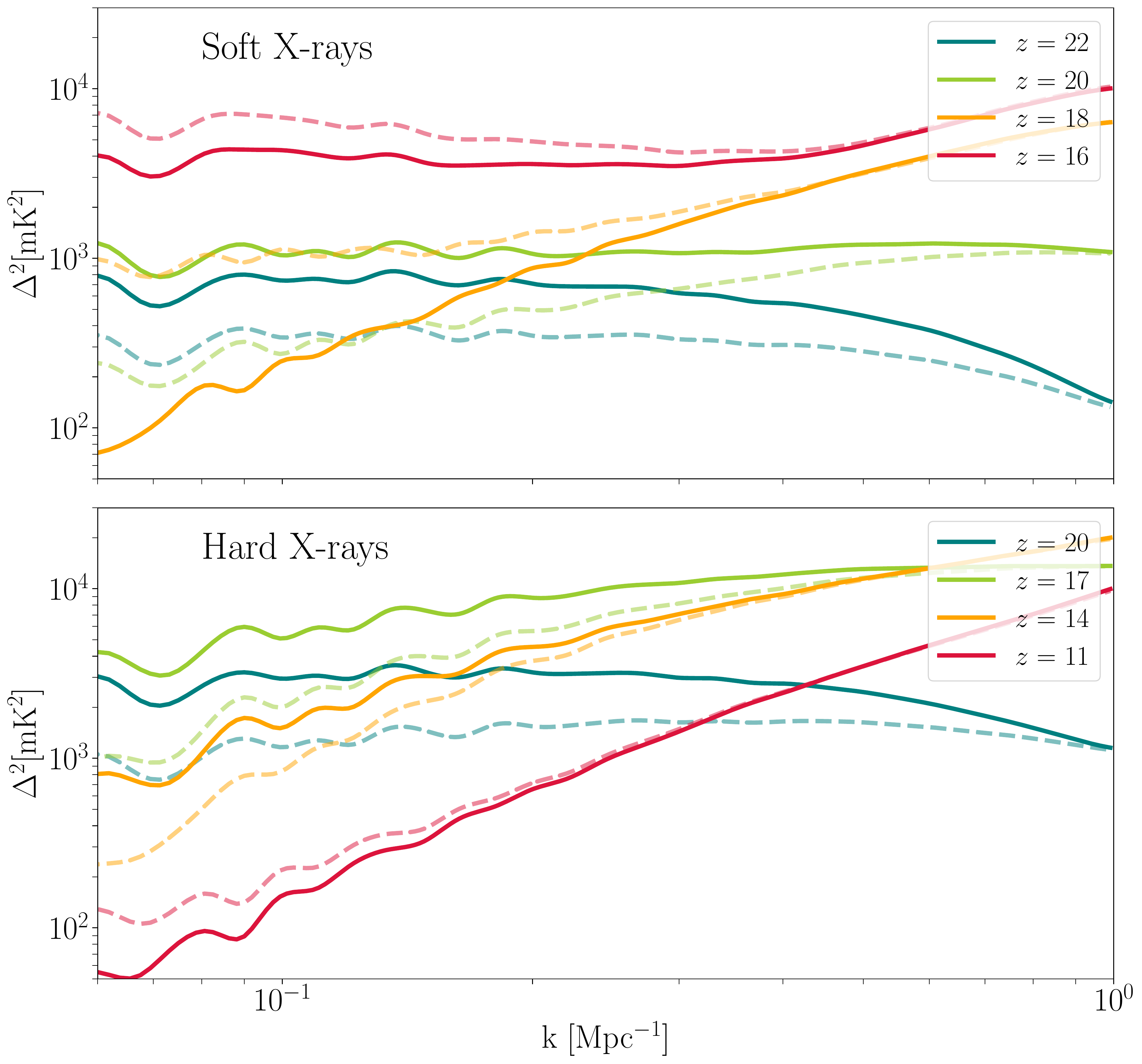}

    \caption{The evolution of the power spectrum shape. Solid lines show the power spectrum with fluctuations in the radio background, while dashed lines show the power spectrum for a uniform radio background with the same mean intensity at each redshift. We show this for the same $f_{\rm Radio} = 3000$ models as in Fig.~\ref{fig:power_spectrum_vs_f_radio}. We show the power spectrum for a different set of redshifts in the two panels since the signature of the radio fluctuations is seen until a later time for the hard X-rays model.}
    \label{fig:ps_shape}
\end{figure}

It is also interesting to look at the shape of the power spectrum at a given redshift. Fig.~\ref{fig:ps_shape} shows the effect of the radio fluctuations on the power spectrum shape for the $f_{\rm Radio} = 3000$ models from Fig.~\ref{fig:power_spectrum_vs_f_radio}. As above, the radio fluctuations enhance the 21-cm fluctuations at high redshifts (when \Lya{} fluctuations dominates), suppress them later on (when X-ray heating fluctuations dominate, which occurs at lower redshifts for the hard X-ray spectrum), and have a reduced effect as we approach the EoR. In this figure we can see that the effect of the radio fluctuations is strongly concentrated on large scales (low $k$), since the radio waves reach far away from the sources, as they are limited only by time-retardation (unlike \Lya{} photons, which have a strict horizon given by atomic frequency ratios, and X-ray photons, which are attenuated by absorption, particularly for a soft X-ray spectrum). Thus, when radio fluctuations enhance the power spectrum they give it a redder slope, and when they suppress the power spectrum they give it a bluer slope. These are potentially observable signatures of radio fluctuations.

\section{The effect of an excess radio background on the 21-cm signal for a standard population of  radio-galaxies}
\label{sec:weak_radio}

Previous works exploring the effects of an astrophysical  radio background on the 21-cm signal \citep{mirocha19, ewall20}, were driven by the EDGES discovery and, thus, focused on  models with extremely bright radio sources, far above what is expected based on extrapolating from observations of low-redshift galaxies. In order to create a deep 21-cm absorption signal, such as observed by EDGES, the sources need to be far brighter in the radio band than present-day galaxies (see the next section for details). Even though a population of weaker radio sources (more similar to the present-day galaxies with $f_{\rm Radio}=1$) cannot account for the deep EDGES signal, it is useful to explore the implications of a realistic population of radio sources on the 21-cm signal in case the EDGES detection turns out to be of a non-cosmological origin \citep[e.g.,][]{Hills:2018, Bradley:2019, Singh:2019, Spinelli:2019,Sims:2020}.  In this section we investigate signatures of radio sources in the 21-cm signal for a wide range of values of $f_{\rm Radio}$, focusing on moderately enhanced backgrounds.

\begin{figure}
    \centering
 \includegraphics[width=0.49\textwidth]{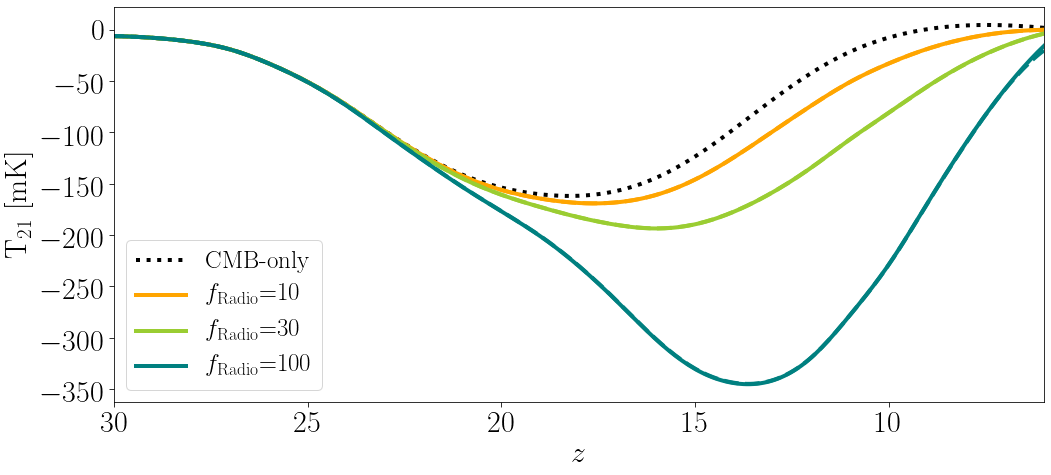}  \\
 \includegraphics[width=0.49\textwidth]{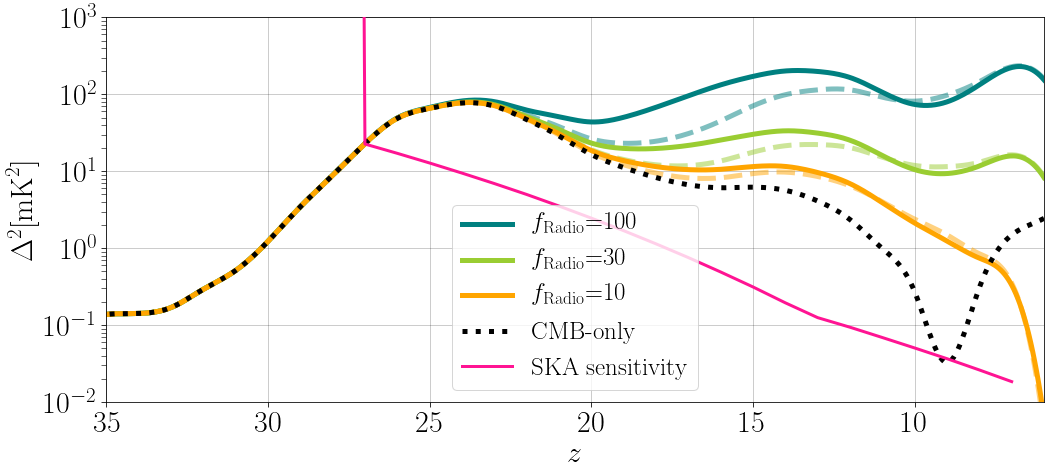} 
 \includegraphics[width=0.49\textwidth]{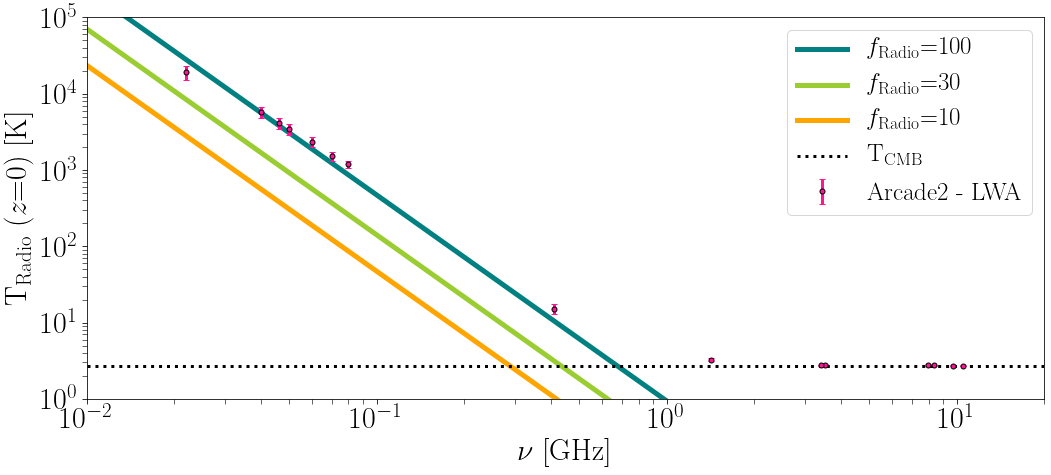}
    \caption{Effect of models with a moderate radio background on
    the 21-cm signal for \Vc{} = 4.2 km s$^{-1}$,  $f_* = 0.1$ and $f_{X} = 0$ and with $f_{\rm Radio}$ = 10, 30, or 100 (as indicated in the legends).  {\bf Top:}  The global 21-cm signal. {\bf Middle:} The 21-cm power spectrum at $k = 0.1~$Mpc$^{-1}$ as a function of redshift. Here we show the case of the fluctuating background (solid) and the corresponding smooth radio background case (dashed) as in the previous section. We also show the projected SKA1 sensitivity (magenta) and the CMB-only case (dotted).     {\bf Bottom:} The present-day excess radio background (relative  to the CMB level shown with the horizontal dotted line). Note that the contribution of galaxies to the radio background is integrated over time from the beginning of the simulation to its end at $z = 6$. The data points  correspond to the excess radio background detected by the ARCADE2 and LWA1 experiments.  
    }
    \label{fig:low_fr_effect}
\end{figure}

As we have seen in the previous section, the  radio-induced fluctuations in the 21-cm signal  anti-correlate with the effect of X-rays and positively correlate with the fluctuations sourced by the \Lya{} and density fields. Therefore, the net effect depends not only on the  properties of the excess radio background, but also on the parameters regulating  star formation ($f_*$ and $V_c$)  and the X-ray luminosity of sources ($f_X$ and the X-ray SED). In comparison with a reference case where the background radiation is the CMB,  the strongest signature of the radio background is expected in  models with negligible X-ray heating accompanied by  early and efficient star formation (i.e., low $V_c$ and high $f_*$). Note that even for a negligible contribution of X-rays, gas in our models is heated up by \Lya{} photons \citep[in agreement with][]{chuzhoy07, ghara19}. This effect sets a heating floor,  creates a plateau in the 21-cm power spectrum (compared to a peak in the case of heating by soft X-rays), and (to some extent) counteracts the effects of an enhanced radio background. In Fig.~\ref{fig:low_fr_effect} we demonstrate the  effect of an excess radio background on the global 21-cm signal and the power spectrum in models with moderate values of the radio production efficiencies ($f_{\rm Radio}$=10, 30 and 100) and for fixed values of $V_c = 4.2$ km s$^{-1}$,  $f_* =0.1$ and no X-ray heating (i.e., $f_X = 0$). 

The maximum normalized difference,
\begin{equation}
\label{eqn:normalized_diff}
    {\rm Normalized\; difference} = \frac{\left|T_{21}^{\rm CMB-only} - T_{21}^{\rm With\;radio}\right|}{{\rm max}\left|T_{21}^{\rm CMB-only}\right|}
\end{equation}
 calculated over the redshift range of $z = 30-6$, in the depth of the absorption trough  is of  19\%, 56\%, and 171\% for $f_{\rm Radio}$=10, 30 and 100, respectively. 
For all cases the maximum increase (i.e., the maximum value of $\left|T_{21}^{\rm CMB-only} - T_{21}^{\rm With\;radio}\right|$)  occurred at $z \sim 12$.
 In addition, while in the CMB-only case the 21-cm signal is seen in emission at low redshifts (even without X-ray heating), in models with moderate  $f_{\rm Radio}$  the signal is in absorption for the entire duration of the EoR.

The effect of a moderate population of radio galaxies on the 21-cm power spectrum is also significant in these models and is of order 5\%, 23\%, and 294\% for $f_{\rm Radio}$=10, 30 and 100, respectively. For all cases the maximum difference occurs at $z \sim 13$. The difference is calculated as in Eq.~\ref{eqn:normalized_diff}, using the power spectrum $\Delta^2$ (at $k = 0.1$ Mpc$^{-1}$) instead of $T_{21}$. We note that using the maximum over all redshifts in the denominator gives a conservative number, i.e., at some redshifts the relative difference is much larger. Because it takes time for the radio background to build up, the strongest effect is at the intermediate and low-redshift parts of cosmic dawn ($z\lesssim 20$) which are the main target of the existing and upcoming interferometers (including AARTFAAC Cosmic Explorer,  LWA, HERA, NenuFAR, and the SKA). At the EoR redshifts, the signal is strongly enhanced due to the presence of the extra radio background, and, because there is no heating transition in this case, does not go through a null; if realised in nature, this is good news for the 21-cm experiments.

We also estimate the contribution of such a population of radio sources to the  present day excess radio background at the frequencies observed by experiments such as ARCADE2 and LWA1 (bottom panel in Fig.~\ref{fig:low_fr_effect}). Here we assume that the early population of galaxies with enhanced radio emission does not survive until today (e.g., due to the redshift evolution of their intrinsic properties) and impose a cutoff redshift of $z = 6$ (or higher, see below). In all the explored cases the contribution of the high redshift radio galaxies is at or below  the background level  detected by ARCADE2 and LWA1, which serves as an upper limit for such a contribution.

\begin{figure}
    \centering
 \includegraphics[width=0.49\textwidth]{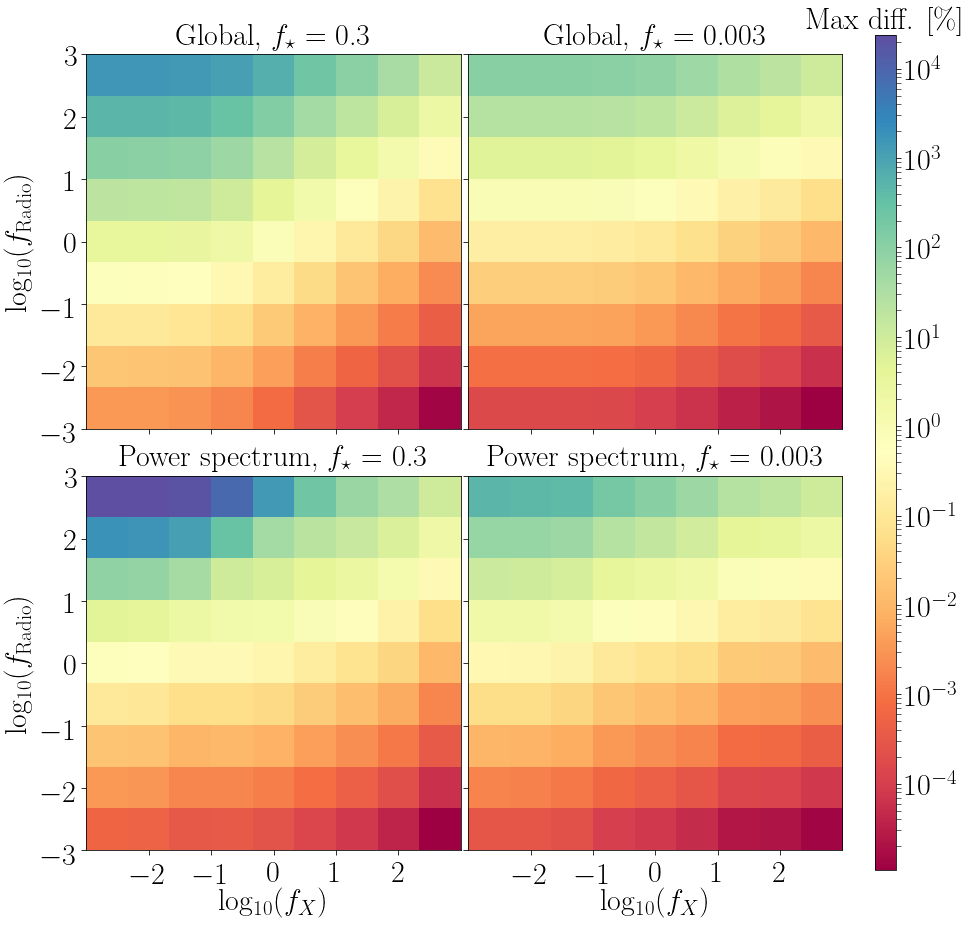} 
    \caption{The maximum normalized difference (in \%)  in the global signal ({\bf top panels}) and power spectrum at $k = 0.1$ Mpc$^{-1}$ ({\bf bottom panels})  between a model with an excess radio background (with f$_{\rm Radio}$ shown on the $y$-axis) and a model with no excess radio background ($f_{\rm Radio} = 0$), for various values of $f_X$ (as indicated on the $x$-axis). We show the results for efficient ($f_* = 0.3$, {\bf left}) or inefficient ($f_* = 0.003$, {\bf right}) star formation and assuming $V_c = $ 4.2 km s$^{-1}$, an X-ray population with a hard SED \citep[using an X-ray binary SED as in][]{fialkov14a}, and reionization parameters of $\zeta = 11.4$ and R$_{\rm mfp}$ = 30 Mpc.}
    \label{fig:low_fr_diff}
\end{figure}

Next, we explore the effect of a moderate radio background in a variety of astrophysical scenarios in a more systematic way (Fig.~\ref{fig:low_fr_diff}). We conduct a parameter study in which we vary  $f_{\rm Radio}$ and $f_X$ on a grid, assuming efficient ($f_* = 0.3$) or inefficient ($f_* = 0.003$) star formation and assuming $V_c = $ 4.2 km s$^{-1}$. Here we assume a population of hard X-ray sources  \citep[using an X-ray binary SED as in][]{fialkov14a}. We note that for this analysis we switch off Poisson fluctuations in the number of galaxies in order to avoid random noise in the results. We calculate the maximum  difference (in absolute value) in both the global signal (top panels of Fig.  \ref{fig:low_fr_diff}) and the power spectrum at $k = 0.1$ Mpc$^{-1}$ (bottom panels of Fig.  \ref{fig:low_fr_diff}), between a case with an excess radio background and the same case with no excess radio background ($f_{\rm Radio} = 0$, i.e., the CMB-only case). We normalize by the maximum value, again in absolute value, of the no-radio signal: for the global signal, we normalize by the maximum absorption depth (Eq.~\ref{eqn:normalized_diff}), and for the power spectrum we normalize by the maximal $\Delta^2$ at $k = 0.1$ Mpc$^{-1}$ (Eq.~\ref{eqn:normalized_diff}, with $\Delta^2$ instead of $T_{21}$). The maximum difference is again found over the redshift range of $z = 30-6$. The result is shown in Fig.~\ref{fig:low_fr_diff}.

We find that there is some difference between models with the excess background and with $f_{\rm Radio} = 0$ (for otherwise identical parameters) for all the explored cases. As anticipated, models with inefficient X-ray production are significantly more sensitive to an excess radio background. In such models we can see a potentially measurable effect (up to $10 \%$) in both the global signal and the power spectrum even when the high redshift galaxy population is assumed to have similar radio brightness to the present day galaxies ($f_{\rm Radio} \sim 1$). On the other hand, for models with efficient X-ray heating, a moderate excess radio background has only a small effect. This happens because  in such cases X-rays quickly raise the kinetic temperature significantly above that of the total radio background and the heating term in Eq.~\ref{eqn:tb} saturates. For low X-ray efficiencies, models with a higher SFR density (SFRD) at high redshifts, i.e., with higher $f_*$ and lower $V_c$, are more sensitive to the effect of the excess radio background. 

\section{Extreme Scenarios for  EDGES Low-Band}
\label{sec:parameters}

\subsection{21-cm observables}

We now use our model that simulates radio emission arising from galaxies (as described in Section~\ref{sec:radiobck}) to explore the relatively extreme cases  that could explain the anomalously deep  EDGES low-band signal \citep{bowman18}. We explain below the differences between our study and that previously done by \citet{mirocha19}. The analysis here is similar to the one performed by \citet{fialkov19} for a phenomenological homogeneous  excess radio background model with a synchrotron power spectrum (the external radio model). As was mentioned in \cref{sec:radiobck}, compared to the external radio model, radio from galaxies is non-uniform and features a different time evolution (increasing with time, as opposed to decaying as in the other case). 
The effect of the radio fluctuations on the global signal is minor and  the main difference between the two models (in the context  of the parameter constraints with EDGES low-band) is due to  the different redshift evolution. Of course, the new  fluctuations do affect  the corresponding power spectra.

To compare the two scenarios, we generated two new datasets of models\footnote{Even though an analysis of external radio background models was done by \citet{fialkov19}, we re-generated the dataset for this paper because of the revised astrophysical model as discussed in \cref{sec:methods}. Specifically, \Lya{} and radiative heating significantly affect the depth of the absorption trough in the scenarios with negligible X-ray heating. The revised  predictions for the external radio background model are discussed in Appendix \ref{sec:param_constraints_Ar}. } in which we systematically varied the astrophysical parameters over the following  ranges: $f_* = 0.01 - 0.5$, $V_{c}$ = 4.2 - 50 km s$^{-1}$,  $f_{X} =0.01-1000$, $\nu_{\rm min}$ = 0.1 - 3 keV,  $\alpha$ = 1 - 1.5, R$_{\rm mfp}$ = 10 - 70 Mpc, and $\tau = 0.03 - 0.089$ in agreement with CMB Planck measurements \citep{planckcollaboration18}. The radio background efficiency was varied in the  $f_{\rm Radio}$ = 10 - 10$^{6}$ range. We created $\sim 10158$ models with radio from galaxies.  We also create $5077$  external radio background models with a radio background strength $A_{\rm r}$ = 0.2 - 1000.

Models were tagged to be EDGES-compatible if they satisfied the following criteria \citep{fialkov18, fialkov19}:
 \begin{multline}
 \label{eqn:econ1}
     300 {\rm mK} < \rm{max}[T_{21} (60 < \nu/{\rm MHz} < 68)] - \\ 
                     \rm{min}[T_{21} (68 < \nu/{\rm MHz} < 88)] < 1000 {\rm mK}, 
 \end{multline}
and 
 \begin{multline}
  \label{eqn:econ2}
     300 {\rm mK} < \rm{max}[T_{21} (88 < \nu/{\rm MHz} < 96)] - \\ 
                     \rm{min}[T_{21} (68 < \nu/{\rm MHz} < 88)] < 1000 {\rm mK}, 
 \end{multline}
 corresponding to the signal detected in the EDGES low-band data \citep{bowman18}.

As was discussed by \citet{mirocha19, ewall20},  the EDGES-motivated models require extremely strong radio sources with $f_{\rm radio} \sim 1000$ at high redshifts. If such a population persisted until lower redshifts, these sources would overproduce the diffuse radio background observed by ARCADE2 and LWA1. As we have seen in the previous section (Fig.~\ref{fig:low_fr_effect}), even a moderate population of radio galaxies with $f_{\rm radio} \sim 100$  had to be truncated at $z=6$ in order not to saturate the observed radio background. Therefore, we allow for the possibility that the enhanced radio emission was a strictly high-redshift phenomena, and introduce a radio cutoff redshift $z_{\rm cutoff}$, in order to ensure consistency with the present-day excess radio background. For each model we found the lowest $z_{\rm cutoff}$ for which the predicted present-day excess radio background does not exceed the observations.

\begin{figure}
    \centering
 \includegraphics[width=0.49\textwidth]{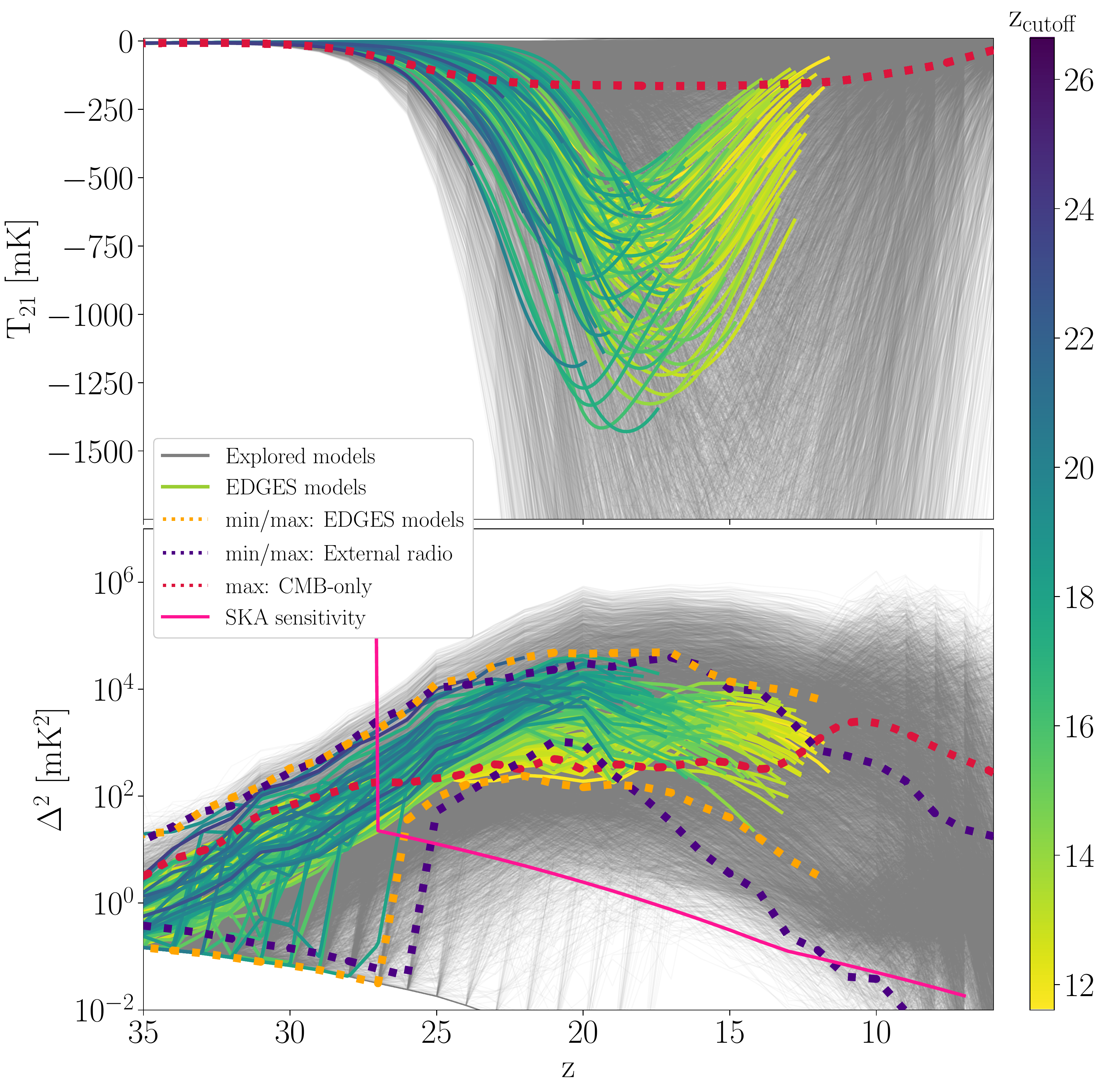} 
    \caption{The global 21-cm signal ({\bf top panel}) and power spectrum at k = 0.1 Mpc$^{-1}$ ({\bf bottom panel}) obtained in our exploration of the parameter space. A random subset of the models that are compatible with the EDGES measurement are shown in color, with the envelope of all such models shown with dotted orange lines ("min/max: EDGES models" in the legend). A broader range of models (not necessarily consistent with EDGES) is shown in grey. The color of each EDGES-compatible model  represents the redshift at which the radio emission needs to be truncated in order not to over-predict the ARCADE2/LWA1 measurement of the present-day extragalactic radio background (the colors are indicated in the colorbar on the right, and each curve is cut off in the plot at the corresponding redshift). We also show the SKA1 sensitivity (pink),  the  maximum envelopes of cases with no excess radio background ($f_{\rm Radio} = 0$,  dotted red), and max and min envelopes of the EDGES-compatible models for the external radio background  (purple dotted line, "min/max: External radio" in the legend). }
    \label{fig:edges_models}
\end{figure}

 In Fig.~\ref{fig:edges_models} we show the global signals and power spectra of the models that are compatible with the  EDGES low-band data. For comparison we also show an envelope (min and max signals at every $z$) of the EDGES-compatible external radio models as well as the maximum signal at every $z$  of the CMB-only cases (calculated using the same updated astrophysical framework as the other two datasets).  We find that the power spectra in the radio from galaxies model can exceed the standard CMB-only cases by two orders of magnitude at $z\sim 20$, thus being a good target for low-band interferometers. At lower redshifts, the power spectra of the EDGES-motivated models are typically  lower compared to the standard models \citep[and compared to excess radio models that are not constrained by EDGES;][]{Mondal:2020}. This is because EDGES requires the absorption signal to be relatively narrow, deep and localized at high redshifts (and thus not at lower redshifts). This also constrains the astrophysical parameter space, in particular the high-redshift star formation history \citep[in a similar way as found by][for the external radio background model]{fialkov19}.

The measured high-redshift absorption signal features a rapid drop into absorption between $z \sim 22$ and $z\sim 17$. Such an early and rapid transition requires the WF coupling to be efficient already at $z\sim 20$, implying substantial star formation at even higher redshifts \citep{fialkov19, Schauer:2019}. This is possible only in scenarios with relatively low values of $V_c$ in which the first stars form in halos of $\sim 10^8$ M$_\odot$ or lower. Moreover, the need for an efficient \Lya{} coupling sets a lower limit on $f_*$. Using the parameter study, we find a lower limit of $f_* = 0.035$ and an upper limit of $V_c  = 35$ km s$^{-1}$. In addition, constrained by the high required \Lya{} intensity, the lower limit on $f_*$ is higher for higher values of $V_c$ (but only above the atomic cooling threshold of 16.5~km/s, since the feedback effects noted in Section~2 limit the contribution of halos below this value). These constraints can be read off parameter plots like Fig.~\ref{fig:parameter_space} (see also the similar plots for other parameter combinations in the Appendix, Figs.~\ref{fig:edges_models_Ar} and \ref{fig:edges_models_FR}).

\begin{figure}
    \centering
 \includegraphics[width=0.49\textwidth]{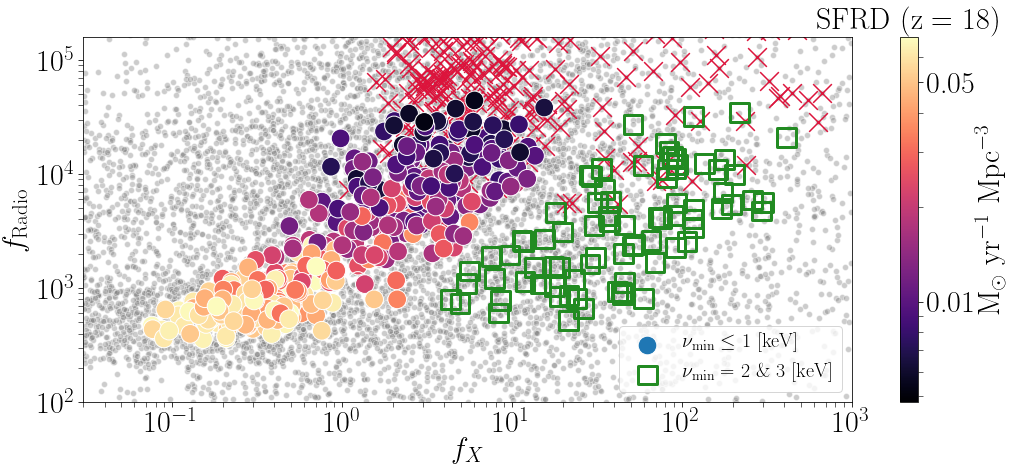} \\ \includegraphics[width=0.49\textwidth]{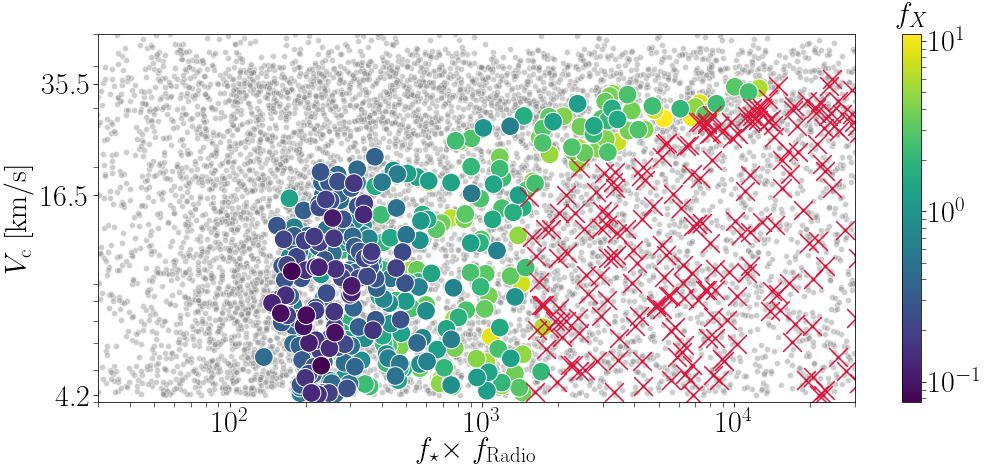}
    \caption{The astrophysical parameters of models compatible with the global 21-cm EDGES low-band measurement  and ARCADE2/LWA1 measurement of the present day excess radio background. Colored markers (circles and squares) show the compatible models. The grey dots show all the explored models.  {\bf Top panel:} $f_{\rm Radio}$  vs.\ $f_X$. We separate the compatible models into two groups. The main group (colored circles) shows models with relatively soft X-ray sources ($\nu_{\rm min}\leq 1$ keV) color-coded with respect to the values of SFRD at $z=18$ (as shown in the colorbar to the right of the panel).
    The second group contains cases with  hard X-ray SEDs  ($\nu_{\rm min}=2$ or 3 keV). Note that the gap between the two groups is due to the sparsity of $\nu_{\rm min}$  values used  in the parameter study.
    {\bf Bottom panel:} $V_c$  vs.\ the total effective radio production efficiency ($f_{*} \times f_{\rm Radio}$). The compatible models are color-coded with respect to the value of $f_X$ (indicated on the colorbar, right-hand side). Red crosses  show models that are compatible with EDGES but not with ARCADE2/LWA1.}
    \label{fig:parameter_space}
\end{figure}

In the standard CMB-only case the depth of the absorption feature is set by the competition between the efficiency of the coupling and the heating. In such models, the EDGES detection cannot be explained as, with the \Lya{} heating taken into account, we find\footnote{This is based on a parameter study of 5077 models with the CMB as the background radiation that we conducted after revising the astrophysical framework and that we will further discuss elsewhere.} a deepest absorption of 164~mK.
When excess radio background is added to the picture, it increases the contrast between the background and the spin temperature, thus deepening the absorption trough. The depth of the observed feature, therefore, sets a lower limit on the possible  radio production efficiency. In the models with the radio background from galaxies it is the combination of $f_* \times f_{\rm Radio}$  (which we refer to as the total effective radio production efficiency) that regulates the number of produced radio photons and, thus, the depth of the absorption trough. 
We find the lower threshold on the total effective radio production efficiency to be $136$  (bottom panel of Fig.~\ref{fig:parameter_space}). In terms of $f_{\rm Radio}$, the lower limit is $356$ , i.e., the high redshift galaxies have to be more than 350 times brighter in radio  than present-day galaxies. This is consistent with \citet{mirocha19}, who found a best fitting value for the radio production efficiency of $f_{\rm Radio} \sim 1000$.  For the lowest $f_{\rm Radio}$ models, the observed deep absorption can be achieved only if the star formation efficiency is high (or, equivalently, the SFRD is high, top panel of Fig.~\ref{fig:parameter_space}).

An additional constraint on the value of $f_* \times f_{\rm Radio}$  comes from the  ARCADE2/LWA1 measurements of the present day radio background. Thus, in Fig.~\ref{fig:parameter_space} and in the rest of this paper, we selected as "EDGES compatible" only models for which $z_{\rm cutoff} < 16$ to ensure that this artificial radio cutoff occurs after most of the era probed by EDGES (we leave for the future a more detailed study of gradual cutoffs). This condition sets an upper limit on $f_* \times f_{\rm Radio}$ , which we find to be $1740$  for models with low $V_c$ (bottom panel of Fig.~\ref{fig:parameter_space}). The threshold is higher for higher $V_c$ models, since in these models galaxies start forming later and, even with the maximum allowed value of star formation efficiency $f_* = 50\%$, the SFRD can be lower than for models with lower $V_c$ in which typical galaxies are smaller but more numerous. Therefore, in the models with high $V_c$ it is possible to have a higher radio production efficiency without exceeding the  ARCADE2/LWA1 measurements.

While \Lya{} photons couple the spin temperature to the temperature of the gas and, thus, create the absorption signal, X-ray photons help to shape the absorption feature. In the standard scenarios, with the CMB being the radio background, it is the total intensity of soft X-rays that regulates the location of the minimum of the absorption 21-cm signal and the steepness of the high-frequency side of the trough. When an excess radio background is added to the picture, it contributes to the structure of the absorption trough (as the signal approximately scales as $1-T_{\rm rad}/T_K$). Therefore, the intensity and redshift evolution of the radio background can strongly affect the shape of the global signal. We find that, to explain the narrow trough detected by EGDES, a fine-tuned contribution of the X-ray population is required, which sets both a lower and an upper limit on $f_X$; these depend strongly on the X-ray spectrum (top panel of Fig.~\ref{fig:parameter_space}) as only photons below $\sim 1$ keV can efficiently heat up the gas. Specifically, for soft sources with $\nu_{\rm min}\leq1$ keV we find $0.08\lesssim f_{ X} \lesssim  10 $, while for hard sources with  $\nu_{\rm min}=2$ or 3~keV the allowed $f_X$ range shifts to higher values ($3\lesssim f_{ X} \lesssim  1000$)\footnote{Note that  $f_X=1000$ is the limit of the prior range set on $f_X$. For higher values of $f_X$, it is likely that we over-predict the unresolved X-ray background observed by {\it Chandra X-ray Observatory} \citep{Fialkov:2017}.}. As expected, we also find tight correlations among $f_{\rm Radio}$, $f_{X}$ and the SFRD (at a given redshift). Models with low $f_{\rm Radio}$ require low $f_{X}$ and a high SFRD to create a deep and narrow global signal, while models with high $f_{\rm Radio}$ require high $f_{X}$ and a low SFRD to create a similar feature.

Using a similar model as explored here but without radio fluctuations, \citet{mirocha19} also reported a lower limit on 
$f_X$, although of a higher value $f_{X} = 10$. Two major differences between our constraint and the result of \citet{mirocha19} are:
\begin{enumerate}
    \item \citet{mirocha19} did not include \Lya{} heating and used a relatively hard X-ray SED \citep[taken from][]{mitsuda84} that is similar to the SEDs of our hard X-ray models shown with green squares in Fig.~\ref{fig:parameter_space}. In contrast, we do include \Lya{} and radiative heating and also explore soft X-ray SEDs; the latter implies more efficient heating and permits lower values of $f_X$. 
 \item     Another difference between our work and \citet{mirocha19} is the different prescription for star formation efficiency. While we use a mass-independent $f_*$ above the atomic cooling threshold (and a logarithmic suppression below), \citet{mirocha19} assume a mass-dependent $f_*$ in which lower-mass halos have significantly lower $f_*$ values, and thus, lower values of the SFRD. 
 Since, as discussed above, only low-mass halos are abundant at $z\sim17$, their low SFRD needs to be compensated by high values of $f_X$ to create the EDGES feature. They also allowed redshift dependence in $f_*$.
\end{enumerate}
We also note that comparing our results to \citet{mirocha19} is not straight forward, since these authors seem to have neglected the  effect of the excess radio background on the coupling coefficients. That is, in Eqs. \ref{eqn:xalpha} and \ref{eqn:col}, they apparently used $T_{\rm CMB}$ instead of $T_{\rm rad} = T_{\rm CMB} + T_{\rm Radio}$\footnote{See the discussion below their Eq.~2.}.  They did not consider the effect of the excess radio background on radiative heating of the IGM (that is, neglected the heating given by Eq.~\ref{eqn:epsilon_rad}). Finally, at least some of the differences between our results and the results of \citet{mirocha19} are explained by the fact that each work applies a different set  of conditions to decide which models fit the EDGES data. Because the total width of the feature can be comparable to the bandwidth of EDGES Low, and the overall zero-point is unknown (due to the foreground removal), here we used a relative criterion (i.e., checking that the changes in the theoretical signals within the EDGES band are compatible with the detection). However, \citet{mirocha19} used an absolute measure (i.e., checking that the maximum absorption agrees with what EDGES have reported). This difference means that we allow for deeper absolute signals (which have less heating) compared to the ones selected by \citet{mirocha19}.

Finally we note that our constraints on the value of $f_X$ are qualitatively different from those derived by \citet{fialkov19} for the external radio background model, where arbitrarily low values of $f_{ X}$ were allowed. This fundamental difference is caused by the different redshift dependence of the radio background, which increases with time in the models considered here and decreases in the case of the external radio background with spectral index $-2.6$ (matching Galactic synchrotron) as considered by \citet{fialkov19}. An increasing radio background acts to deepen the absorption at lower redshifts and, thus, X-rays need to compensate for this effect in order to produce the narrow trough seen by EDGES. On the other hand, a decreasing radio background helps to produce a narrow absorption feature even with a very weak or negligible X-ray contribution. Additional differences between the two radio background models are discussed in Appendix \ref{sec:param_constraints_Ar}.

\subsection{Radio source counts}

In scenarios with high values of $f_* \times f_{\rm Radio}$ , such as the models that are compatible with the EDGES low-band signal, high-redshift radio galaxies might be bright enough to be directly seen by radio surveys \citep[e.g., see][]{Nitu:2020}. 
Here we explore which of the EDGES-compatible models predict bright radio galaxies that are expected to be above the detection thresholds of existing and upcoming surveys. 

\begin{figure}
    \centering
 \includegraphics[width=0.49\textwidth]{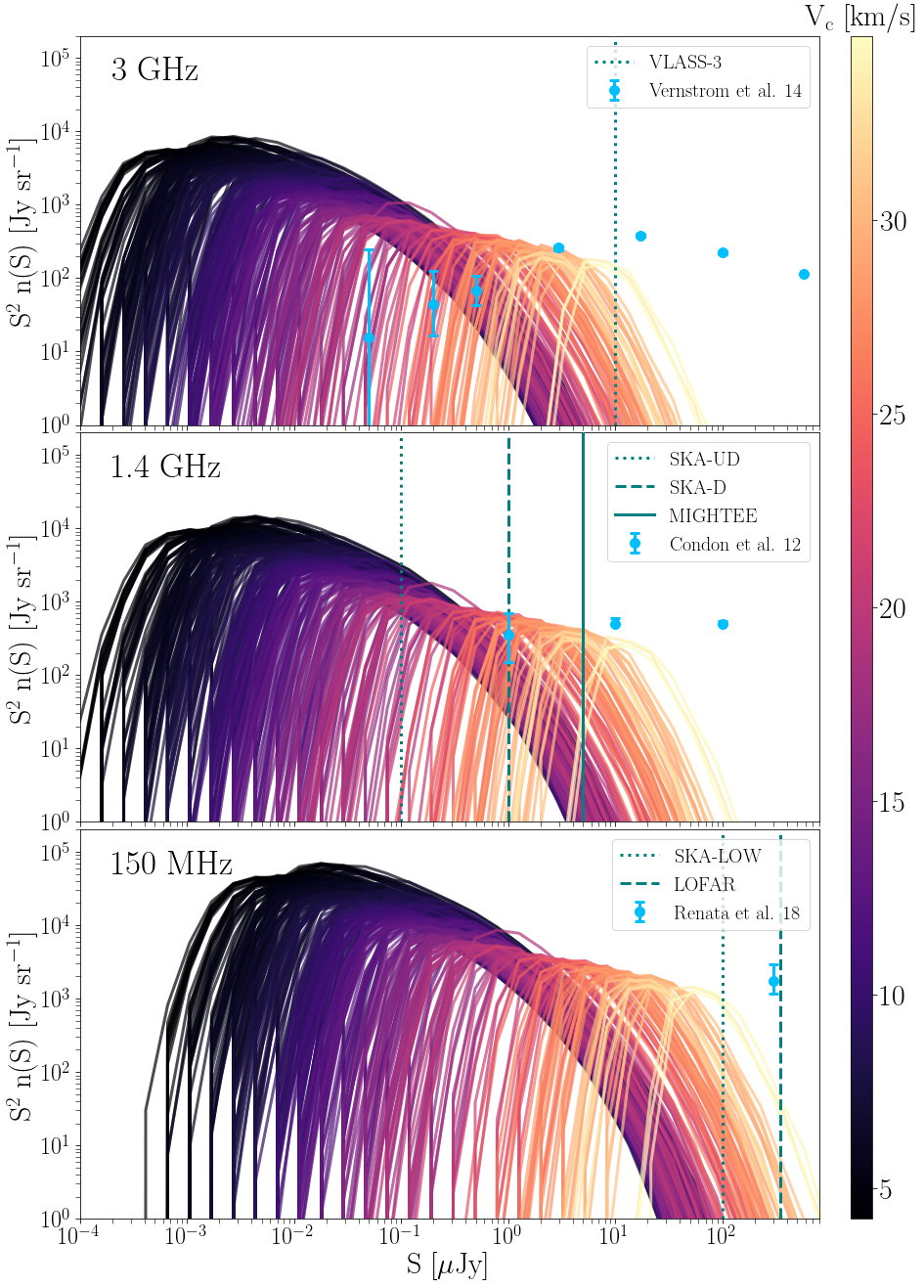} 
    \caption{Radio source counts for our EDGES-compatible models at 3 GHz ({\bf top}), 1.4 GHz ({\bf middle}) and 150 MHz ({\bf bottom}). Each curve is color-coded with respect to the values of $V_c$ (as indicated in the color-bar on the right). In each panel blue points show current observational constraints with 1-$\sigma$ error bars from \citet{vernstrom14}, \citet{condon12}, or  \citep{retana18}. Vertical lines show  5-$\sigma$ detection thresholds for future and current surveys (see legends for details). }
    \label{fig:source_counts}
\end{figure}

An individual galaxy of intrinsic radio luminosity $L_{\nu}$,  calculated using Eq.~\ref{eqn:eps_radio} from the SFR of the galaxy, can be detected if the corresponding observed flux density $S_{\nu}$
\begin{equation}
    S_{\nu} = (1+z) \frac{L_{\nu  (1+z)}}{4 \pi D^2_L(z)}\ , 
\end{equation}
where $D_L$ is the luminosity distance, is above the sensitivity limit of a telescope. We calculated $n(S_{\nu})$, the number of sources per steradian per unit flux density, expected for each model. As in the previous section, for each model we do not include redshifts below $z_{\rm cutoff}$ (the redshift where the enhanced radio needs to be truncated in order to not exceed the ARCADE2/LWA1 measurements). At each redshift we find the number density of galaxies that produce a given observed flux density in the range of $S_{\nu}$ to $S_{\nu} +dS$, from which we calculate the number per steradian. We divide by dS to get the number per steradian per unit flux. Finally, we integrate over all redshifts to get the total $n(S_{\nu})$.
The predictions (in the commonly used combination of $n(S_{\nu}) S_{\nu}^2$) for flux densities at $\nu = 3$ GHz,  $\nu = 1.4$ GHz, and  $\nu = 150$ MHz, are shown in Fig.~\ref{fig:source_counts}.  We compare our results  to current constraints from   fluctuation analysis of deep radio images taken with the Karl G.\ Jansky Very Large Array (VLA) at 3 GHz \citep{vernstrom14} and at 1.4 GHz \citep{condon12}, as well as to an  individual 5-$\sigma$ source detection at 150 MHz \citep{retana18} using LOFAR. In addition we show 5-$\sigma$ detection thresholds for current and future surveys including MIGHTEE \citep{jarvis16}, LOFAR \citep{shimwell19}, VLASS-3 and  SKA \citep{prandoni15}.

The source count distributions shown in Fig.~\ref{fig:source_counts} are color-coded with respect to the values of $V_c$ for each model. As expected, the source counts  are dominated by faint sources, well below the various sensitivity thresholds, for models with low values of $V_c$. In such scenarios star formation starts especially early (owing to the hierarchical nature of structure formation) and radio galaxies are typically small but numerous. On the other hand, the luminosity functions of  high $V_c$ models are dominated by bright but scarce objects. We treat the observed points as upper limits on our models, since we do not include the contribution of low-redshift galaxies (below the cutoff redshift for each model). In some of these models the number counts  exceed  the observational constraints at 3 GHz, and models with $V_c \sim 25$~km/s are disfavored at a significance of a few $\sigma$. Comparing the model predictions to the expected detection thresholds of the future surveys with MIGHTEE \citep{jarvis16}, VLASS-3 and  SKA \citep{prandoni15}, we see that such observations can be used to further constrain the subset of EDGES-compatible models presented in this work.

\section{Summary}
\label{sec:summary}

In this paper we considered three  aspects of a high-redshift population of radio galaxies:

First, our main novelty is that we examined the effect of fluctuations in the radio background produced by a population of high-redshift radio galaxies on the resulting 21-cm power spectrum. We found that the radio background fluctuations affect the 21-cm signal at redshifts and scales targeted by existing and future interferometers, such as HERA and the SKA. The 21-cm power spectrum is enhanced at the onset of star formation owing to the positive correlation between the effects of radio, the density field, and \Lya{} coupling on the 21-cm signal. The 21-cm  power spectrum is suppressed by the effect of radio fluctuations at later redshifts when the heating contribution is important, due to the negative correlation between the effects of the radio background and heating on the 21-cm signal. The strength of both the suppression and the enhancement of the power spectrum are model-dependent. For example, we found qualitatively different signatures in models with soft and hard X-ray SEDs. Also, locally, the radio emission from galaxies enhances the absorption signal around individual sources during cosmic dawn. This effect should be important for 21-cm imaging.

Next, we explored the implications, for both the global signal and the power spectrum, of a normal population of radio galaxies, i.e., galaxies with radio efficiencies only moderately enhanced over the present-day population. We quantified the effect of the excess radio background, varying astrophysical parameters such as the minimum mass of star-forming halos, the X-ray heating efficiency, star formation efficiency and the amplitude of the radio background. We found that in models with a low X-ray efficiency and high star formation rate (which depends on both the circular velocity and star formation efficiency), the radio background from standard radio galaxies  can have an effect of a few tens of percent on the 21-cm signal. Therefore, it is important to account for the potential radio contribution in an accurate 21-cm analysis.

Finally, we performed an exploration of the astrophysical parameter space to identify models that are compatible with the tentative EDGES low-band detection. We found that high redshift galaxies need to be at least $ 350$ times more efficient in producing low-frequency radio waves compared to present-day galaxies in order to produce a sufficient radio background to explain the EDGES detection. Comparing our models  to the reported absorption feature, we found a lower limit of $f_*=0.035$, and an upper limit of $V_c = 35$ km s$^{-1}$. We also found that the X-ray heating efficiency is restricted to a relatively narrow SED-dependent range. E.g., for soft sources with $\nu_{\rm min}\leq1$ keV the limits are  $0.08\lesssim f_X\lesssim 10$, and the range shifts to higher values of $f_X$ for harder SEDs. We compared these limits to the prediction of a model where the excess radio background is smooth and has a synchrotron spectrum (the external radio background model). The major difference between the two possible models is in the time evolution of the radio background (increasing with time in the former case versus decaying in the later). This difference results in there being no limits on $f_X$ in the external radio background model. A major implication of the  enhanced radio background is that it boosts the 21-cm power spectrum by a few orders of magnitude compared to the standard case where the background radiation is the CMB. Such an enhanced signal might be within the sensitivity of AARTFAAC Cosmic Explorer, NenuFAR, LWA, and HERA.
Finally, we computed the expected radio source counts for the EDGES-motivated models and compared them to the existing observations of LOFAR and VLA as well as to the detection threshold of future radio surveys. We found that existing observations can already significantly constrain some of the model parameters. It will certainly be exciting to see the results of upcoming observations of radio sources as well as new 21-cm measurements.

\section*{Acknowledgments}

We acknowledge the usage of the DiRAC HPC. AF was supported by the Royal Society University Research Fellowship. This project was made possible for I.R.\ and R.B.\ through the support of the ISF-NSFC joint research program (grant No.\ 2580/17). 

This research made use of:
 {\fontfamily{cmtt}\selectfont  SciPy} \citep[including {\fontfamily{cmtt}\selectfont  pandas} and {\fontfamily{cmtt}\selectfont NumPy, }][]{2020SciPy, numpy},  {\fontfamily{cmtt}\selectfont IPython} \citep[][]{perez07}, {\fontfamily{cmtt}\selectfont matplotlib} \citep[][]{hunter07}, {\fontfamily{cmtt}\selectfont Astropy} \citep[][]{astropy-collaboration13}, {\fontfamily{cmtt}\selectfont Numba} \citep{lam15},  the SIMBAD database \citep[][]{wenger00}, and the NASA Astrophysics Data System Bibliographic Services.

\section*{Data availability}
No new data were generated or analysed in support of this research.

\bibliographystyle{mnras}
\bibliography{cosmo}

\appendix

\section{Effect of an excess radio background on the global 21-cm signal}
\label{sec:radio_general_effect_example}
\begin{figure}
    \centering
 \includegraphics[width=0.49\textwidth]{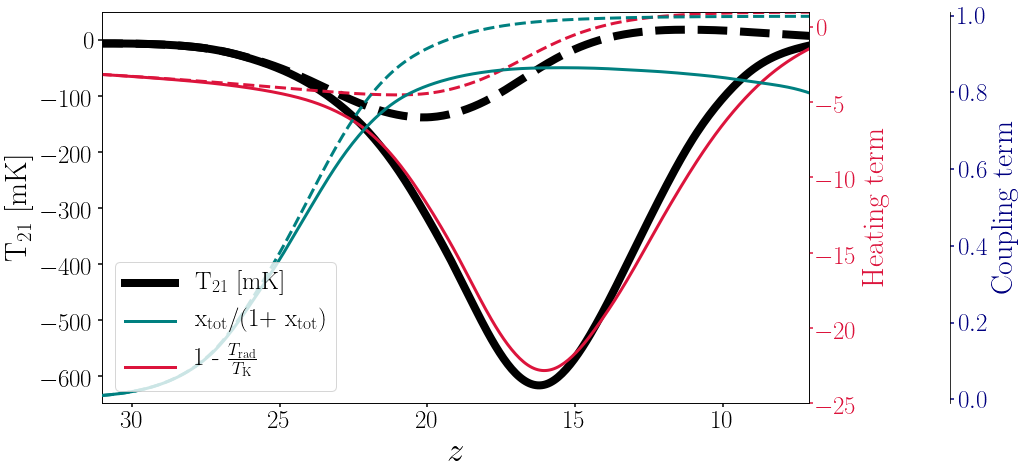}      \\

    \caption{The effect of the excess radio background on the 21-cm signal. We show the same astrophysical model with (solid lines, $f_{\rm Radio} = 1000$) or without (dashed lines, $f_{\rm Radio} = 0$) an excess radio background. The other astrophysical parameters are the same as for the hard X-ray example of Figs. \ref{fig:power_spectrum_vs_f_radio} and \ref{fig:ps_comps}. This example shows that while the dominant effect is due to the heating term, the effect due to the coupling term is also significant. }
    \label{fig:means}
\end{figure}

In this section we give further details on the effect of an excess radio background on the global 21-cm signal, for the model where the radio background is produced by galaxies. We begin with an example separating out two different effects. We use the same astrophysical parameters as for the hard X-ray example of Figs.~\ref{fig:power_spectrum_vs_f_radio} and \ref{fig:ps_comps}. Fig.~\ref{fig:means} shows the evolution of the 21-cm global signal, the coupling term, and the heating term, with ($f_{\rm Radio} = 1000$) or without ($f_{\rm Radio} = 0$) an excess radio background.

As discussed in \cref{sec:radio_back}, the excess radio background affects the 21-cm signal in a number of ways, enhancing the contrast between the spin temperature and the background radiation temperature, decreasing the coupling coefficients, and enhancing the radiative heating. While the enhanced contrast between the spin temperature and $T_{\rm rad}$ is the dominant effect, in this example we can see that the reduced coupling term is also important, producing a $\sim 20\%$ effect. The coupling term is reduced from $\sim 1$ to $\sim 0.8$ after the coupling transition, due to the excess radio background. The reason that the coupling term remains roughly constant (after the initial rise) in the model with excess radio background is that both the \Lya{} and excess radio emission are proportional to the SFR, so that, to a first approximation, the redshift dependence cancels out in the coupling coefficient (which is $\propto J_{\alpha}/T_{\rm rad}$, see \cref{sec:radio_back}). We note that in this example, heating is dominated by X-rays, so the effect of the excess radio background on radiative heating is negligible.

\begin{figure}
    \centering
 \includegraphics[width=0.49\textwidth]{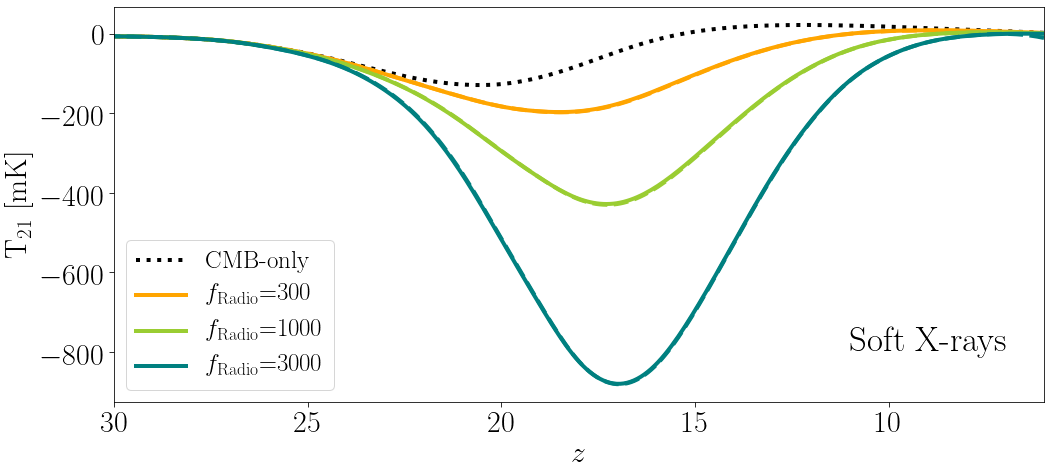} \\  
 \includegraphics[width=0.49\textwidth]{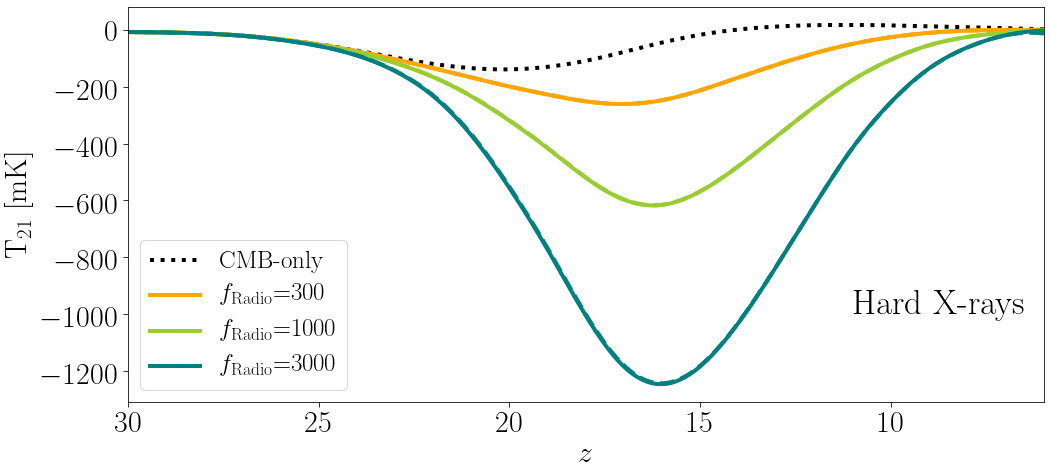}
    \caption{The 21-cm global signal for different values of $f_{\rm Radio}$ (as indicated in the legend) and for  two different X-ray SEDs: soft ($\nu_{\rm min} = 0.1$ keV, {\bf top}) and hard ($\nu_{\rm min} = 1$ keV, {\bf bottom}).   Values of  the other parameters are fixed (and match Fig.~\ref{fig:power_spectrum_vs_f_radio}): $V_c = 16.5$ km s$^{-1}$, $f_*$ = 0.1, $f_{X}$ = 1, $\alpha = 1$.  We show cases with the full, fluctuating radio background from galaxies (solid) and with the corresponding smooth radio background with the same mean radio intensity at every $z$ as in the fluctuating case  (dashed). Note that in this case the dashed lines are very similar to the solid lines, meaning radio fluctuations do not have a significant effect on the global signal. We also show the case with no excess radio background (i.e., the CMB-only case, black dotted line). }
    \label{fig:global_vs_f_radio}
\end{figure}

Next we show, in Fig.~\ref{fig:global_vs_f_radio}, the 21-cm global signal for various models with high values of $f_{\rm Radio}$. This is the global signal version of Fig.~\ref{fig:power_spectrum_vs_f_radio}. Radio fluctuations only have a small effect on the global signal, so the comparison case (of a smooth radio background with the same mean intensity at each redshift) is nearly indistinguishable from the full model in each case. 
Both the strength of the radio background and the X-ray spectrum strongly affect the global 21-cm signal.

\section{Full set of parameter constraints using EDGES low-band}
\label{sec:param_constraints_Ar}

In this section we discuss the similarities and differences between the parameter constraints (\cref{sec:parameters}) for the two excess radio background models discussed in this paper: the main model where the radio background is created by galaxies, and the external radio background with a synchrotron radio spectrum as explored by  \citet{fialkov19}.

In order to perform a direct comparison between the radio from galaxies and the external radio background models, we had to re-run the models of \citet{fialkov19} using the upgraded-physics simulation employed here (\cref{sec:methods}). Compared to  \citet{fialkov19}, the updated simulation included Poisson fluctuations, multiple scattering, and \Lya{} heating (\cref{sec:methods}).   We obtained $5077$  updated models with the external radio background,  530 of which agree with EDGES.

\begin{figure*}
    \centering
 \includegraphics[width=0.3\textwidth]{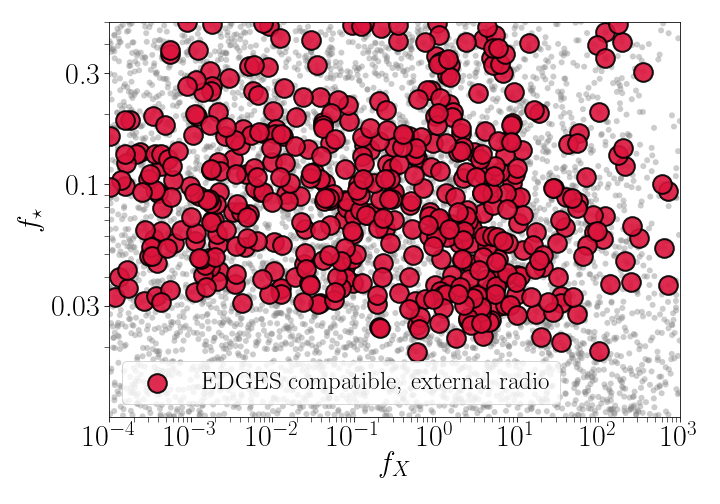}\includegraphics[width=0.3\textwidth]{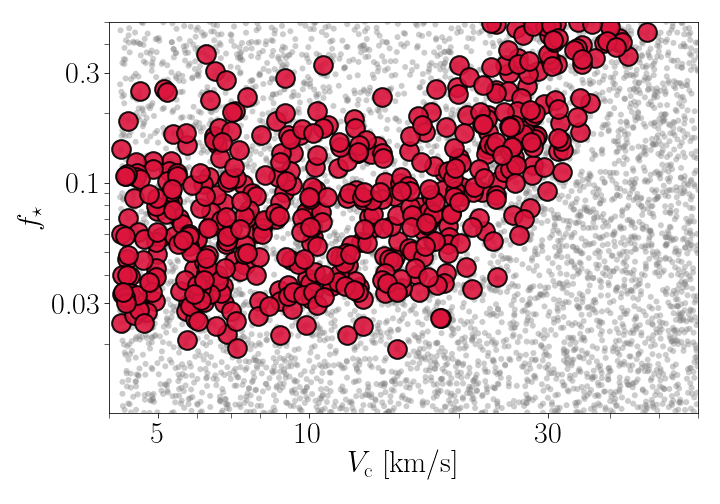}\includegraphics[width=0.3\textwidth]{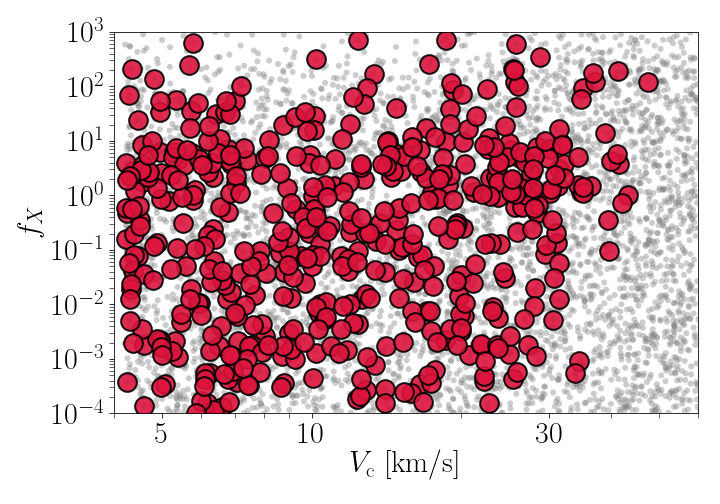} 
 \includegraphics[width=0.3\textwidth]{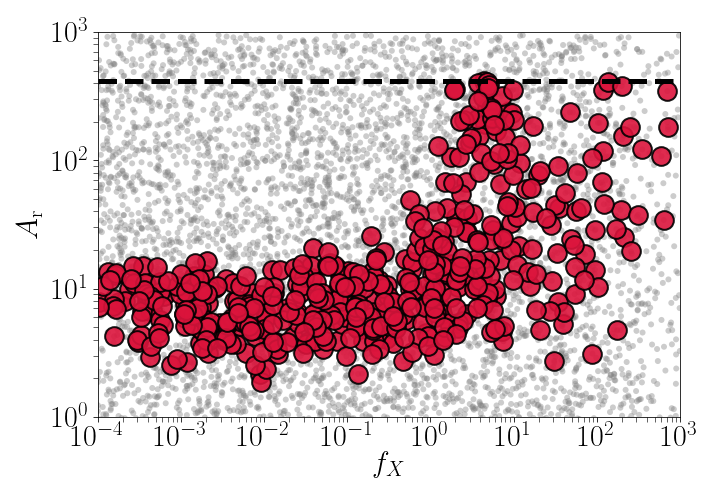}\includegraphics[width=0.3\textwidth]{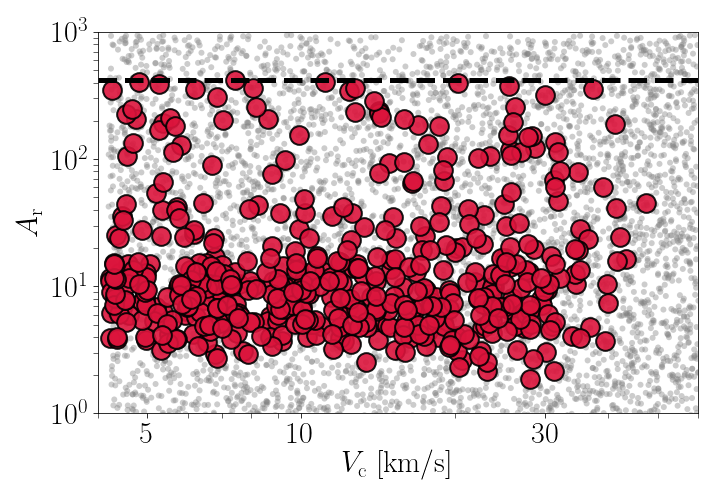}\includegraphics[width=0.3\textwidth]{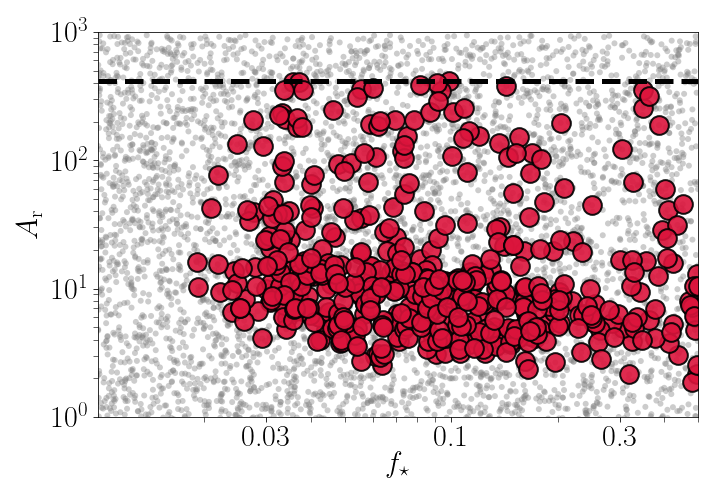} 
    \caption{EDGES low-band parameter constraints for the external radio model. Models explored are shown in grey, and models compatible with EDGES low-band are shown in color. The horizontal dashed lines show the upper limit on $A_{\rm r}$ derived from ARCADE2/LWA1 measurements.}
    \label{fig:edges_models_Ar}
\end{figure*}

\begin{figure*}
    \centering
 \includegraphics[width=0.3\textwidth]{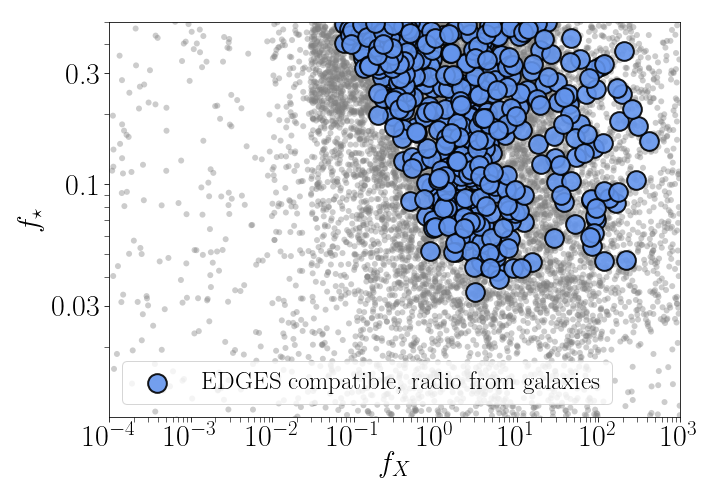}\includegraphics[width=0.3\textwidth]{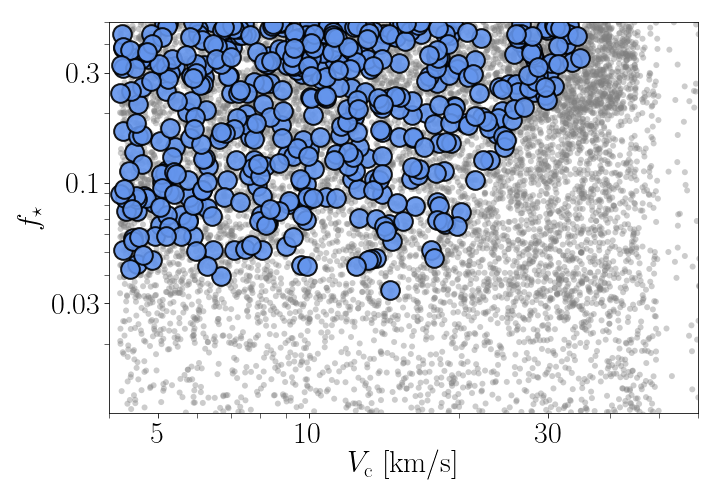}\includegraphics[width=0.3\textwidth]{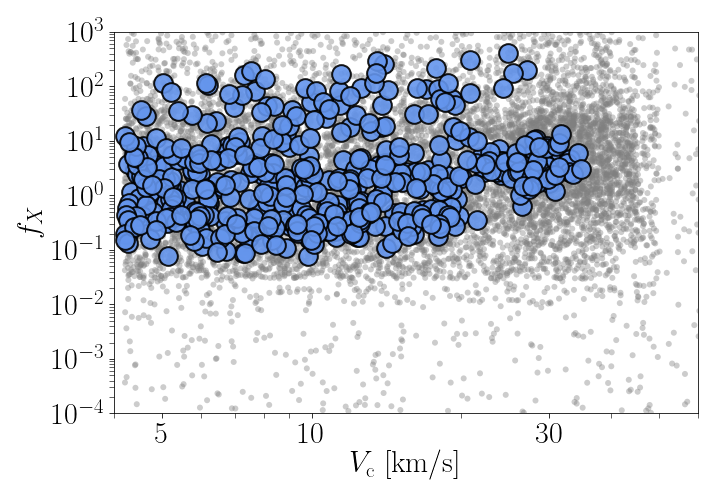} 
 \includegraphics[width=0.3\textwidth]{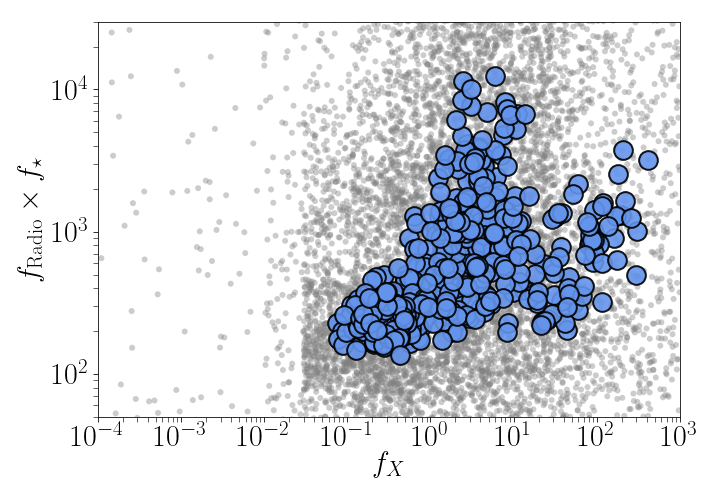}\includegraphics[width=0.3\textwidth]{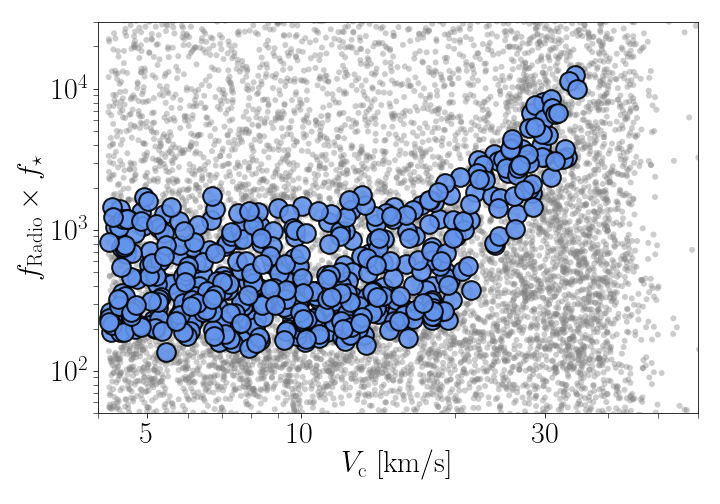}\includegraphics[width=0.3\textwidth]{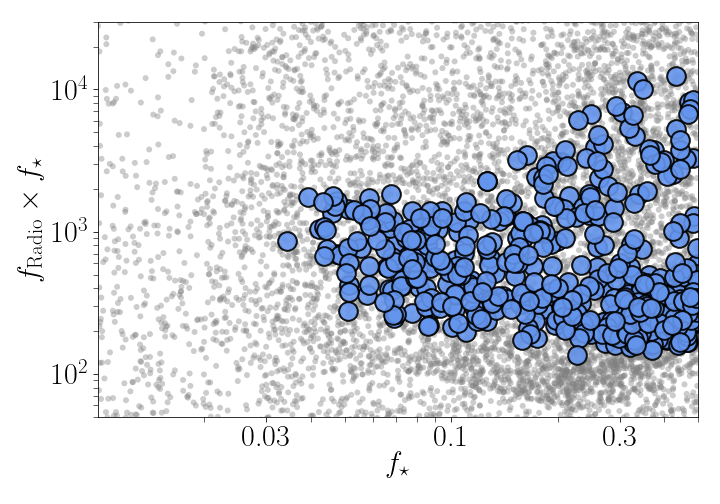} 
    \caption{EDGES low-band parameter constraints for the radio from galaxies model. The explored parameters are shown in grey, and models compatible with EDGES low-band are shown in color. }
    \label{fig:edges_models_FR}
\end{figure*}

Figs. \ref{fig:edges_models_Ar} and \ref{fig:edges_models_FR} summarize  the astrophysical constraints of the EDGES compatible models for the two radio background scenarios. The biggest qualitative difference between the two models is the different redshift dependence: in the model with a radio background from galaxies the radio background builds up with time as it traces the star formation activity, and its intensity is greater at lower redshifts; while in the external radio background case the intensity is lower at lower redshifts. This difference is reflected in the astrophysical constraints (predominantly on $f_X$): 
\begin{enumerate}
    \item As discussed in the main text, there is no constraint on $f_{X}$ in the external radio background case, while for radio from galaxies we find an upper and lower limit for soft X-ray sources ($\nu_{\rm min} \leq1$ keV) of $0.08\lesssim f_X\lesssim 10$. For harder X-ray sources these limits shift to higher values of $f_X$ as such sources are intrinsically  less efficient in heating up the gas. 
\item For the radio from galaxies model, lower values of $f_{X}$ are allowed at higher $f_*$. This is because the overall heating rate (which is proportional to both $f_*$ and $f_{X}$) has to be above some threshold value to fit the narrow absorption trough observed by EDGES. On the other hand, for the external radio background model, there is no lower limit on $f_{X}$ for any value of $f_*$.
\item  For a given $f_*$, the highest allowed $V_c$ is slightly lower for the radio from galaxies model, compared to the external radio background model. 
\item The lowest allowed $f_*=0.035$ is higher for the radio from galaxies model, compared to the  external radio background model, where it is $f_* = 0.019$.
\item The upper limit on $V_c = 35$  km s$^{-1}$ is lower for the radio from galaxies  model, compared to the  external radio model, where it is $V_c = 47.5$ km s$^{-1}$.  
\item The minimum allowed value on the radio background parameter is   $A_{\rm r} = 1.9$  for the external background, and $f_{\rm Radio} \times f_* \sim 140$ or $f_{\rm Radio} \sim 360$ for the other model. It is not possible to directly compare these quantities as they are defined quite differently.  
\end{enumerate}

% Don't change these lines
\bsp	% typesetting comment
\label{lastpage}
\end{document}